\documentclass[acmsmall,screen]{acmart}
\setcopyright{none}
\renewcommand\footnotetextcopyrightpermission[1]{}
\makeatletter\def\nocountry{\global\@ACM@countrypresenttrue}\makeatother

\usepackage{cspsymb}
\usepackage{simplebnf}
\usepackage{mathpartir}
\usepackage{float}
\usepackage{booktabs}
\usepackage{algorithm}
\usepackage{algpseudocode}
\newfloat{listing}{tbp}{lol}
\floatname{listing}{Listing}
\usepackage{xcolor}
\usepackage{tikz}
\usetikzlibrary{arrows.meta,positioning,shapes.geometric,automata}
\usepackage{listings}
\lstdefinelanguage{CoreFi}{
  morekeywords={MemoryRegion, ProtectionDomain, Chan, System,
                Mapping, Events, Event, Process, import, as, Low, High},
  sensitive=true,
  morecomment=[l]{//},
  morestring=[b]",
}
\newcommand{\ltrans}[1]{\xrightarrow{#1}}
\newcommand{\btrans}[1]{\xRightarrow{#1}}


\newcommand{\terminated}{\mathit{term}}

\title{CoreFi: A Core Calculus for Embedded Systems with Bounded Shared-Memory Channels}

\author{Yan Liu}
\affiliation{%
  \institution{Australian National University, Australia\\ \normalfont yan.liu@anu.edu.au}
  \nocountry}
\author{Alex Potanin}
\affiliation{%
  \institution{Australian National University, Australia\\ \normalfont alex.potanin@anu.edu.au}
  \nocountry}
\authorsaddresses{}



\begin{document}

\begin{abstract}
Embedded systems are built from components that compute on private state and communicate over bounded buffers in shared memory; since a component must keep running rather than wait on a slower one, a sender drops a message when the buffer is full, and a receiver blocks when it is empty.
Process calculi, the established basis for reasoning about communicating systems, model these channels as bounded queues that block a sender when full.
The code itself, written in a low-level language such as C, intertwines computation and communication so tightly that the message-passing protocol is hard to recover automatically for verification.
We present CoreFi, a core calculus for embedded systems with bounded shared-memory channels, derived from Communicating Sequential Processes (CSP), that addresses both.
Our channel semantics is faithful to embedded practice, and we extend CSP's abstract events to concrete computation events, calls to foreign functions, so that a component's program in CoreFi becomes an orchestration of its communication and computation events.
The composition of the component processes is defined by the scheduling logic of the underlying kernel, one rule that says when a component runs its body next, from an initial configuration the boot leaves, rather than by a parallel operator; we instantiate it on the verified microkernel seL4 and its MicroKit framework.
Deadlock freedom and determinism are two fundamental properties of embedded systems, and we prove both for every well-formed system, that is, one whose communication dependencies are acyclic and whose guarded choices partition the state space.
The separation of computation from communication also separates local from global state change: computation events, which change only a component's local state, can be absorbed into big steps, and the abstracted components compose to a smaller model of the global behaviour; a refinement relation guarantees that trace properties proved on the abstracted global state space hold of the complete one.
We provide the formalisation and the proofs in Lean~4, assisted by a coding agent.
\end{abstract}

\maketitle
\pagestyle{plain}


\section{Introduction}\label{sec:intro}

Embedded systems run network appliances and protocol stacks, driver-assistance functions in cars, flight control in aircraft, and the control systems of critical infrastructure such as power grids and water treatment, where a failure costs more than a crash of the software; many run for years without a restart, so a fault cannot simply be patched after deployment.
Whatever the domain, such a system is built the same way: as a set of components---drivers, protocol handlers, controllers---each running on private state, invoked by a kernel or a scheduler when something arrives for it, and communicating with the others through shared memory.
A network driver, for example, is woken by its device and reads a frame's header to decide whether to forward the frame, reply to it or discard it.
Forwarding copies the frame into a buffer shared with the next component; when that buffer is full the frame is dropped, since the driver must keep running rather than wait.
In C, the most common implementation language, the manipulation of shared memory, the computation and the signalling of other components are interleaved in one program.
Such code is hard to write correctly and harder to verify: properties of the whole system, deadlock freedom first among them, depend on the communication protocol, which can be recovered from the intertwined code neither easily by hand nor reliably by a tool.
It would therefore be ideal to program the communication protocol apart from the \emph{computation}, by which we mean both a component's logic and its manipulation of memory, so that the protocol becomes a program of its own, an \emph{orchestration} of each component's computation and communication steps.
That program can then serve directly as the model of the protocol: formalising it for a model checker~\cite{holzmann97spin,gibson14fdr,uppaal,sun09pat} becomes an automatic step, and the model evolves with the code, its verification results updated rather than invalidated when the program changes.

P~\cite{desai13p} was among the first languages to take this approach: its components are state machines exchanging events over unbounded queues, compiled to code and explored by a systematic tester.
Synchronous dataflow languages such as Lustre~\cite{halbwachs91lustre} and SCADE do the same for control software: a program is verified as written and compiled by a qualified generator, over a clocked schedule and without shared-memory communication between components.
Mimosa~\cite{huber25mimosa,huber25mimosa-compiler} gives a formal semantics to time-triggered processes over FIFO buffers and compiles them to real-time tasks; Lingua Franca~\cite{lohstroh21lf} makes reactor programs deterministic by construction through logical time; on seL4, Kalas~\cite{pohjola22kalas} compiles a global program to verified code, and Tower~\cite{hickey14tower} generates the communication between Ivory tasks.
Code generation from a verified model~\cite{finn23pgo,hackett25tracelink,sun25automan} goes the other way, from a written model to code, keeping the two in step by construction.
Each formalises the program on a communication model of its own, none on bounded buffers that drop when full, and in each the formalisation is either not automated or leaves out the platform's scheduler.

We present CoreFi, a core calculus that takes this approach on the platform's own model, derived from Communicating Sequential Processes (CSP)~\cite{hoare1978communicating}.\footnote{In memory of Sir Tony Hoare (1934--2026), inventor of CSP and Turing Award laureate.}
A component's orchestration is a process term that sequences \emph{computation events}, named blocks that call local code, with sends and receives on channels, under guards over the component's store; the computation stays in the language it is already written in.
For the program to be the model of the system, the semantics must be the platform's: asynchronous channels are bounded, a send to a full buffer dropping its value rather than blocking, synchronous channels are rendezvous, computation is not silent, and a kernel decides when a component runs.
We instantiate the calculus on the verified microkernel seL4~\cite{klein09sel4} and its MicroKit framework~\cite{weibel24}, whose isolated address spaces, channels and memory mappings are the components, channels and regions of the calculus, and whose scheduler supplies the platform's rule.
Our contributions are:
\begin{itemize}
\item \emph{A calculus with no silent steps} (\S\ref{sec:calculus}): a guard fires atomically with the event it guards, since a silent step would resolve a choice invisibly, and termination is a predicate on states rather than an event, so every transition of a component's local labelled transition system (\emph{llts}) carries an observable label, computation events included.
\item \emph{The platform as a rule} (\S\ref{sec:system}): the global labelled transition system (\emph{glts}) composes the \emph{llts}es directly over a shared channel store, reading each component's current \emph{invocation}, a fresh body once the previous run has terminated, from an initial configuration the platform's boot leaves; the process term has no scheduler loop and the composition no parallel operator, and instantiating the calculus on a platform, as we do for seL4 MicroKit, means supplying this rule and that configuration.
\item \emph{Properties by construction} (\S\ref{sec:proofs}): every system keeps its buffers within capacity, and two structural conditions, acyclic dependency and choice partition, make every well-formed system deadlock-free, in an open-system form that tolerates waiting on the environment, and deterministic, each label leading from any reachable configuration to at most one successor, without exploring its state space.
\item \emph{Abstraction beside the composition} (\S\ref{sec:pd-global}): since a computation event changes only its component's local state, an abstraction of each \emph{llts} that absorbs computation into big steps (\emph{alts}) has the same channel traces, and the composition of the abstractions has those of the \emph{glts}, so a trace property checked on the smaller model holds of the system, while determinism, which is not a trace property, is proved of the system itself.
\end{itemize}
All of the above is mechanised in Lean~4~\cite{lean4} with CSLib~\cite{cslib}, developed with the assistance of a coding agent; the development accompanies the paper~\cite{fidus}.
\S\ref{sec:example} introduces the running example, a network data diode, \S\ref{sec:related} discusses related work and \S\ref{sec:conclusion} concludes.

\section{Running Example: Data Diode}\label{sec:example}

A \emph{network data diode} is a hardware-enforced security device that allows data to flow in only one direction---from an untrusted public network to a trusted private network---preventing any reverse channel.\footnote{\url{https://owlcyberdefense.com/products/data-diode-products/}}
We build a software one, and we build it twice: in C, as embedded developers write it today, and in CoreFi.

The system has three components (Figure~\ref{fig:datadiode}): \textsf{eth\_outer} drives the network card on the public side, \textsf{eth\_inner} the one on the private side, and the diode component \textsf{dd} sits between them.
Two one-way channels connect them, $\mathit{o2d}$ from \textsf{eth\_outer} to \textsf{dd} and $\mathit{d2i}$ from \textsf{dd} to \textsf{eth\_inner}, each a ring of packet slots in a memory region shared by its two ends.
The diode forwards every packet it finds on $\mathit{o2d}$ to $\mathit{d2i}$, and drops it when $\mathit{d2i}$ is full rather than wait, so that a slow private side can never hold up the public one.

\paragraph{The diode in C}
Figure~\ref{fig:diode-side} is the diode component as MicroKit's ethernet example writes it~\cite{microkit_ethernet}, forwarding direction only, debug output elided and identifiers shortened; the example's full sources, the two ethernet drivers included, are public.
The C function \lstinline|notified| runs whenever a channel fires (\S\ref{sec:platform} returns to this platform contract) and walks the upstream ring until it finds a free slot.
For each occupied slot it computes the slot's address from a base address and an index, wraps the index by hand, tests the downstream slot's flag, copies the packet and sets the flag, or drops the packet, and finally clears the upstream flag to hand the slot back.
None of this is the diode's decision; it is the ring protocol, and both ethernet drivers repeat it for each ring they touch.
Nothing in the C states that the flow is one way, either: it is one way because no component happens to write to $\mathit{o2d}$ from the private side.

\paragraph{The diode in CoreFi}
\S\ref{sec:syntax} introduces CoreFi by writing the same component, one construct at a time.
The ring protocol becomes the meaning of two channel events, and the one-way flow becomes a fact of the channel declarations.

\definecolor{pdblue}{HTML}{156082}
\definecolor{floworange}{HTML}{E97132}
\definecolor{guardgreen}{HTML}{2E7D32}

\begin{figure}[t]
\centering
\begin{tikzpicture}[
  pd/.style={draw=pdblue, fill=pdblue, rounded corners=3pt,
             minimum width=1.8cm, minimum height=0.75cm,
             align=center, font=\scriptsize\sffamily\color{white}, text=white},
  net/.style={align=center, font=\small\itshape},
  chan/.style={-Stealth, thick, pdblue},
  bidi/.style={Stealth-Stealth, thick, pdblue!60},
  lbl/.style={font=\footnotesize\ttfamily, midway, above=3pt},
  flowlbl/.style={font=\footnotesize\sffamily, midway, above=2pt, text=floworange},
]
  \node[net] (pub)       {Public\\Network};
  \node[pd,  right=1.0cm of pub]       (eth-outer) {eth\_outer\\{\scriptsize pri\,=\,99}};
  \node[pd,  right=1.5cm of eth-outer] (dd)        {DataDiode\\{\scriptsize pri\,=\,150}};
  \node[pd,  right=1.5cm of dd]        (eth-inner) {eth\_inner\\{\scriptsize pri\,=\,199}};
  \node[net, right=1.0cm of eth-inner] (priv)      {Private\\Network};

  \draw[bidi] (pub.north east) to[out=50,in=130]  (eth-outer.north west);
  \draw[bidi] (eth-outer.south west) to[out=230,in=310] (pub.south east);
  \draw[bidi] (eth-inner.north east) to[out=50,in=130]  (priv.north west);
  \draw[bidi] (priv.south west) to[out=230,in=310] (eth-inner.south east);

  \draw[chan] (eth-outer) -- node[lbl]{$\mathit{o2d}$} (dd);
  \draw[chan] (dd)        -- node[lbl]{$\mathit{d2i}$} (eth-inner);

  \draw[-Stealth, thick, floworange, shorten <=2pt, shorten >=2pt]
    ([yshift=6pt]eth-outer.north) --
    node[flowlbl]{Guarantee One-way Data Flow}
    ([yshift=6pt]eth-inner.north);
\end{tikzpicture}
\caption{The data diode.}
\label{fig:datadiode}
\end{figure}
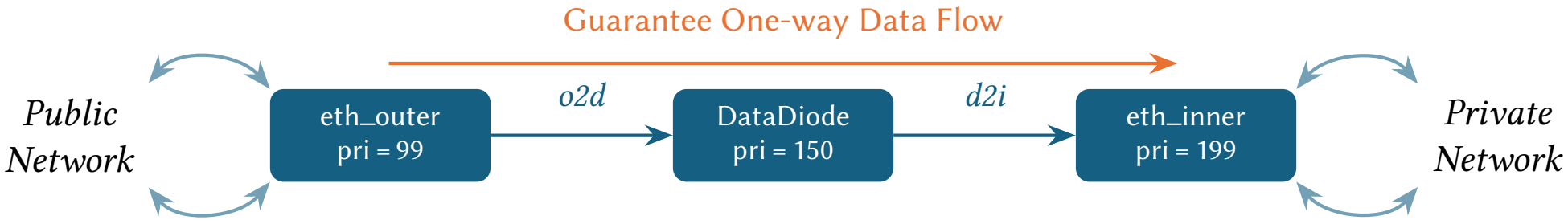

\begin{figure}[t]
\centering
\begin{minipage}[t]{0.40\textwidth}
\begin{lstlisting}[language=C,escapeinside={(*@}{@*)},xleftmargin=1.5em,basicstyle=\scriptsize\ttfamily]
#define BUFFER_SIZE (2 * 1024)(*@\label{c:layout-from}@*)
#define DATA_OFFSET 64
#define BUFFER_MAX  1024
struct buffer_descriptor {
  uint16_t data_length;
  uint16_t flags; };
unsigned in_idx = 0, out_idx = 0;
uintptr_t IN, OUT; /* bases */(*@\label{c:layout-to}@*)
volatile struct buffer_descriptor
  *bd, *obd;
volatile void *pkt, *opkt;

void init(void) {(*@\label{c:init}@*)
  /* debug output only */
}

\end{lstlisting}
\end{minipage}\hfill
\begin{minipage}[t]{0.57\textwidth}
\begin{lstlisting}[language=C,escapeinside={(*@}{@*)},xleftmargin=1.5em,basicstyle=\scriptsize\ttfamily,firstnumber=16]
void notified(microkit_channel ch) {
  switch (ch) {(*@\label{c:switch}@*)
  case OUTER_INPUT_CH:
    for (;;) {(*@\label{c:loop}@*)
      bd  = (void *)(IN + BUFFER_SIZE * in_idx);
      pkt = (void *)bd + DATA_OFFSET;
      if (bd->flags == 0) break;(*@\label{c:empty}@*)
      in_idx = (in_idx + 1) % BUFFER_MAX;(*@\label{c:idx-from}@*)(*@\label{c:idx-to}@*)
      obd = ...; opkt = ...;  /* as above, OUT */
      if (obd->flags == 1) {(*@\label{c:full}@*)
        /* no space downstream: drop */
      } else {
        obd->data_length = bd->data_length;(*@\label{c:send-from}@*)
        mycpy(opkt, pkt, bd->data_length);
        obd->flags = 1;
        microkit_notify(INNER_OUTPUT_CH);
        out_idx = (out_idx + 1) % BUFFER_MAX;(*@\label{c:send-to}@*)
      }
      bd->flags = 0;(*@\label{c:release}@*)
    }
    break;
  }
}
\end{lstlisting}
\end{minipage}
\begin{lstlisting}[language=C,escapeinside={(*@}{@*)},xleftmargin=1.5em,basicstyle=\scriptsize\ttfamily,firstnumber=39]
/* eth.c, the driver: transmit a frame on the NIC's ring (debug output elided) */
static void send_frame(uint8_t *d, unsigned int length)(*@\label{c:send-frame}@*)
{
    uint16_t flags = tbd[tbd_index].flags;
    if (flags & (1 << 15)) { return; }          /* ran out of tx buffers */
    void *packet = (void *)(uintptr_t)(packet_buffer_vaddr
                     + ((RBD_COUNT + tbd_index) * PACKET_BUFFER_SIZE));
    mycpy(packet, d, length);
    seL4_ARM_VSpace_CleanInvalidate_Data(3, (uintptr_t)packet,
                                         ((uintptr_t)packet) + length);
    flags = (1 << 15)   /* ready */
          | (1 << 11)   /* last in frame */
          | (1 << 10);  /* transmit crc */
    if (tbd_index == TBD_COUNT - 1) { flags |= (1 << 13); } /* wrap */
    tbd[tbd_index].data_length = length;
    tbd[tbd_index].flags = flags;
    eth->tdar = (1 << 24);                      /* SEND */
    tbd_index++;
    if (tbd_index == TBD_COUNT) { tbd_index = 0; }
}
\end{lstlisting}
\caption{The diode component in C, and the driver's transmit function.}
\label{fig:diode-side}
\end{figure}

\section{The CoreFi Calculus}\label{sec:calculus}\label{sec:syntax}\label{sec:sos}
A CoreFi program consists of channel declarations and one \emph{component program} per component, which orchestrates the component's local computation and its communication over the channels.
CoreFi is designed to be an \emph{executable} language rather than a declarative specification; following Roscoe's process algebra for CSP~\cite{roscoe2010understanding}, each construct comes with its step-by-step operational semantics, given here beside its syntax.
The process language follows CSP~\cite{hoare1978communicating}, extended with guards over data, assignments and calls into local code, as CSP\#~\cite{sun09pat} and Promela~\cite{holzmann97spin} are; guarded processes and external choices encode conditionals, and recursion encodes loops.
One deliberate departure from vanilla CSP is the omission of internal choice: non-determinism is undesirable in a program, and restricting to external choice is a prerequisite for the determinism result while preserving CSP's compositional reasoning power.

We introduce each construct by writing the diode component with it; the program grows construct by construct.
Programs are written in an ASCII rendering of the abstract syntax: \lstinline|->| for $\then$, \lstinline|[]| for $\extchoice$, and named process definitions for recursion.

\subsection{Channels}\label{sec:channels}\label{sec:state-config}\label{sec:channel-events}
Components communicate only through declared channels.
All CoreFi channels are \emph{single-producer single-consumer}---a well-established practice in embedded systems enabling lock-free, low-overhead, race-free communication---and a channel names its sender and receiver, so that events need no location annotations.

\noindent
\par\vspace{0.3em}\begin{bnf}
$c$ ::= $p$ $\to_m^{N}$ $p$: asynchronous channel with buffer region $m$ and capacity $N \geq 1$
       | $p_1$ $\rightarrowtail$ $p_2$: synchronous channel; no buffer
\end{bnf}\smallskip

Asynchronous channels ($\to_m^N$) are \emph{bounded}: they buffer up to $N$ values in a FIFO, $N_c$ for channel $c$, a send to a full buffer drops its value rather than blocking, and a receive blocks when the buffer is empty---unlike most process models, which assume unbounded channels or block a sender at capacity.
Synchronous channels ($\rightarrowtail$) rendezvous directly, with no buffer.
Every channel runs between two components, except an \emph{interrupt channel}, which wraps a hardware interrupt and has the environment at its other end: a receive on it is the interrupt's delivery, which the environment may send at any time, and a send on it is the acknowledgement, which the environment never refuses, so it never blocks.

\paragraph{Declaration.}
A channel is declared by one of two rules: an asynchronous channel needs two distinct endpoints and a fresh buffer region, which the platform maps into both; a synchronous channel needs only two distinct endpoints, a platform possibly constraining them further:
\begin{mathpar}\scriptsize
\inferrule*[left=\textsc{Chan-Async}]
{c \triangleq (p_1 \to_m^{N} p_2) \qquad c \notin C \qquad p_1 \neq p_2 \qquad N \geq 1 \qquad m \in M \qquad m \notin M^{C}}
{C,\; M^{C} \vdash c \leadsto C \cup \{c\},\; M^{C} \cup \{m\}}
\and
\inferrule*[left=\textsc{Chan-Sync}]
{c \triangleq (p_1 \rightarrowtail p_2) \qquad c \notin C \qquad p_1 \neq p_2}
{C \vdash c \leadsto C \cup \{c\}}
\end{mathpar}

\paragraph{Buffer semantics.}
The buffer of an asynchronous channel $c$ is a FIFO queue $q \in \mathit{Val}^{\leq N_c}$, the front being the next value to leave and the back where the next value arrives.
Two operations define it:
\[
\mathit{enq}(q, v) \;=\; \begin{cases} q \cdot v & |q| < N_c \\ q & |q| = N_c \end{cases}
\qquad\qquad
\mathit{deq}(v \cdot q') \;=\; (v,\, q'),
\]
so an enqueue to a full buffer drops the value and leaves the buffer unchanged, and a dequeue is defined only on a non-empty buffer; a send is an enqueue and a receive a dequeue.
Concretely the platform realises $q$ as a ring of $N_c$ flagged slots over $m$: a write index and a read index advance modulo $N_c$, $\mathit{enq}$ returns at once when the slot at the write index is flagged occupied, and $\mathit{deq}$ clears the flag of the slot at the read index after taking its value.
That ring is what the C of Figure~\ref{fig:diode-side} operates by hand; a CoreFi program never touches it, only sends and receives.
A synchronous channel has no buffer: its send and receive meet in a rendezvous, which is a rule of the composition, not of the component.

\emph{In the diode.}
The diode's two channels are asynchronous, each over a buffer region of the C's \lstinline|buffer_descriptor|s; the capacity is not stated but computed, the region's size over the descriptor's, 1024 for the C's 2\,MiB regions and 2\,KiB descriptors:
\begin{lstlisting}[numbers=none]
chan o2d = eth_outer -> dd         via m_o2d<buffer_descriptor>
chan d2i = dd -> eth_inner         via m_d2i<buffer_descriptor>
\end{lstlisting}
The two lines replace the C's ring layout, its indices and its base addresses (lines~\ref{c:layout-from}--\ref{c:layout-to}), and the system configuration that maps each ring into its two ends.
The direction is part of the declaration: \textsf{dd} can only receive on $\mathit{o2d}$ and only send on $\mathit{d2i}$, and no declaration lets \textsf{eth\_inner} reach \textsf{eth\_outer}.

\subsection{Stores}\label{sec:stores}\label{sec:state}
A component computes on a \emph{local store}, private memory no other component can reach.
What is shared is the channel buffers, which live in memory regions mapped into both endpoints: the system holds each buffer once, in the \emph{global channel store}, and a component holds a \emph{view} of the buffers of its channels.
A local state pairs the process term with the two.

\begin{definition}[Local State Configuration]\label{def:local-store}
The \emph{local state configuration} of component $p$ is the triple
$s^p = \langle \rho^p,\, \sigma^p,\, \gamma^p \rangle$
where:
\begin{itemize}
\item $\rho^p$ is the current process term;
\item $\sigma^p : \mathbb{N} \rightharpoonup \mathit{Val}$ is the \emph{local store}, the component's private memory as a partial map from slot addresses to values, which no other component can reach; and
\item $\gamma^p : C_p \rightharpoonup \mathit{Val}^*$ is the \emph{channel buffer view}, a per-channel FIFO queue over the async channels of $p$, where $C_p = \{\, c \in C_{\mathit{async}} \mid p \in \mathit{sender}(c) \cup \mathit{receiver}(c) \,\}$.
\end{itemize}
\end{definition}

We write $S^p$ for the set of all such triples, the state space of $p$.

\begin{definition}[Global Channel Store]\label{def:global-store}
The \emph{global channel store} $\Gamma : C_{\mathit{async}} \to \mathit{Val}^*$ maps each asynchronous channel $c$ to the finite sequence $\Gamma(c) \in \mathit{Val}^{\leq N_c}$ of values currently held in its FIFO buffer; the front of $\Gamma(c)$ is the next value to be dequeued and the back is where the next enqueue appends.
Synchronous channels have no buffer and are not in the domain of $\Gamma$.
\end{definition}

The view $\gamma^p$ of a component is not a copy: it is read off the store as $\gamma^p = \Gamma|_{C_p}$.
Since all endpoints of $c$ map the same physical region (\emph{Coh}, \S\ref{sec:mappings}), the view of every endpoint agrees with the store:
\begin{equation}
\forall\, c \in C_{\mathit{async}},\; \forall\, p \in \mathit{sender}(c) \cup \mathit{receiver}(c).\quad
\gamma^p(c) \;=\; \Gamma(c)
\tag{G-Coh}\label{inv:gcoh}
\end{equation}
A local rule changes a component's view; the composition writes the changed view back to $\Gamma$, so \eqref{inv:gcoh} is kept by construction.

\subsection{Events}\label{sec:events}
A component program has two parts: \emph{computation events}, named statement blocks of assignments and function calls, and \emph{processes}, which sequence computation events and channel events into the component's execution flow.
The event alphabet of $p$ is $\Sigma^p = \Sigma^p_\alpha \cup \Sigma^p_c$: the computation events $\Sigma^p_\alpha = \{\alpha_1, \alpha_2, \ldots\}$, each fired atomically as one step, and the communication events $\Sigma^p_c = \{\, c!v \mid p \in \mathit{sender}(c),\, v \in \mathit{Val} \,\} \cup \{\, c?v \mid p \in \mathit{receiver}(c),\, v \in \mathit{Val} \,\}$.
A \emph{guarded event} $[b]\,\ell$ is an event $\ell \in \Sigma^p$ annotated with the guard under which it fired (\textsc{Guard-T}); the guard is an execution-time constraint, not a distinct observable, and $[b]\,\ell$ and $\ell$ are indistinguishable in traces.

\paragraph{Notation.}
We write $\llbracket e \rrbracket_{\sigma}$ for the evaluation of expression $e$ in local store $\sigma$, and $s[x \mapsto v]$ for the store $s$ updated to map $x$ to $v$; a program variable $x$ is a slot address of $\sigma$.
For a guard $b$ and a label $\ell \in \Lambda^p$, the \emph{guard merge} $[b]\cdot\ell$ annotates $\ell$ with $b$, flattening nested guards by conjunction:
\[
[b]\cdot\ell = [b]\,\ell \quad \text{if } \ell \in \Sigma^p,
\qquad
[b]\cdot([b']\,\ell_0) = [b \wedge b']\,\ell_0.
\]

\subsubsection{Communication Events}\label{sec:comm-events}
A \emph{channel event} is a single communication action on a declared channel $c \in C$: a send $c!e$ by the channel's sender or a receive $c?x$ by its receiver.
On an asynchronous channel the two are the buffer operations applied to the component's view $\gamma$ (Table~\ref{tab:comm-rules}).
\begin{table}[t]
\scriptsize
\caption{Communication event rules}\label{tab:comm-rules}
\begin{mathpar}
\inferrule*[left=\textsc{Send}]
{c \in \mathit{dom}(\gamma) \quad v = \llbracket e \rrbracket_{\sigma} \quad |\gamma(c)| < N_c}
{\langle c!e \then \rho,\, \sigma,\, \gamma \rangle \ltrans{c!v}
  \langle \rho,\, \sigma,\, \gamma[c \mapsto \gamma(c) \cdot v] \rangle}

\inferrule*[left=\textsc{Loss}]
{c \in \mathit{dom}(\gamma) \quad v = \llbracket e \rrbracket_{\sigma} \quad |\gamma(c)| = N_c}
{\langle c!e \then \rho,\, \sigma,\, \gamma \rangle \ltrans{c!v}
  \langle \rho,\, \sigma,\, \gamma \rangle}

\inferrule*[left=\textsc{Recv}]
{c \in \mathit{dom}(\gamma) \quad \gamma(c) = v \cdot q'}
{\langle c?x \then \rho,\, \sigma,\, \gamma \rangle \ltrans{c?v}
  \langle \rho,\, \sigma[x \mapsto v],\, \gamma[c \mapsto q'] \rangle}
\end{mathpar}
\begin{mathpar}
\inferrule*[left=\textsc{Sync}]
{c \triangleq (p_1 \rightarrowtail p_2) \quad
  s^{p_1} = \langle c!e \then \rho_1,\, \sigma_1,\, \gamma_1 \rangle \quad
  s^{p_2} = \langle c?x \then \rho_2,\, \sigma_2,\, \gamma_2 \rangle \quad
  v = \llbracket e \rrbracket_{\sigma_1}}
{(s^{p_1},\, s^{p_2}) \ltrans{c.v}
  (\langle \rho_1,\, \sigma_1,\, \gamma_1 \rangle,\; \langle \rho_2,\, \sigma_2[x \mapsto v],\, \gamma_2 \rangle)}
\end{mathpar}
\end{table}
\textsc{Send} and \textsc{Loss} are $\mathit{enq}$, with room and at capacity, and \textsc{Recv} is $\mathit{deq}$, binding the head value to~$x$.
On an empty buffer no rule applies: the component has no transition until a send on $c$ fills it, which is the only way a component blocks short of a rendezvous.
On a synchronous channel $c$ from $p_1$ to $p_2$ the two meet in a rendezvous, which is never within one component: \textsc{Sync} relates the pair $(s^{p_1}, s^{p_2})$ of the sender's and the receiver's local states, exchanging the value in one step labelled $c.v$ and touching no buffer; the composition applies it to the two components' current states.
A term contains $c!e$ or $c?x$ only if $p$ is the declared sender or receiver of $c$, so the rules' endpoint assumptions hold by construction.

\subsubsection{Computation Events}\label{sec:comp-events}
A computation event $\alpha$ is a sequence of assignments and external function calls; what it computes is opaque to the calculus, which records only its effect on the local store, possibly none.

\noindent
\par\vspace{0.3em}\begin{bnf}
$\mathit{Block}$ $\alpha$ ::= $x \gets e$ : assignment of expression $e$ to variable $x$
                  | call($g$, $\bar{x}$) : function call with arguments $\bar{x}$
                  | $\alpha$; $\alpha$ : a sequence of assignments and calls
\end{bnf}\smallskip

where $g$ names a function of the external implementation, imported from its header and called by its qualified name, such as \lstinline|eth.send_frame|, and $\bar{x}$ are slot addresses passed by reference.
Expressions $e$ are those of a common expression language over constants and the slots of the local store, and a guard $b$ is a boolean combination of comparisons between expressions; neither calls a function, and neither reads a channel: a component sees a message only by receiving it.

\begin{table}[t]
\scriptsize
\caption{Computation event rules}\label{tab:event-rules}
\begin{mathpar}
\inferrule*[left=\textsc{Evt-Assign}]
{v = \llbracket e \rrbracket_{\sigma}}
{\mathit{effect}(x \gets e,\, \sigma,\, \gamma) = \sigma[x \mapsto v]}

\inferrule*[left=\textsc{Evt-Call}]
{\sigma' = \llbracket g \rrbracket(\bar{x},\, \sigma)}
{\mathit{effect}(\mathit{call}(g,\, \bar{x}),\, \sigma,\, \gamma) = \sigma'}

\inferrule*[left=\textsc{Evt-Seq}]
{\sigma_0 = \sigma \qquad
  \forall\, i \in 1..n.\quad
  \mathit{effect}(\alpha_i,\, \sigma_{i-1},\, \gamma) = \sigma_i}
{\mathit{effect}(\alpha_1 \mathbin{;} \cdots \mathbin{;} \alpha_n,\, \sigma,\, \gamma) = \sigma_n}

\inferrule*[left=\textsc{Evt}]
{\alpha \in \Sigma_\alpha \qquad \mathit{effect}(\alpha,\, \sigma,\, \gamma) = \sigma'}
{\langle \alpha \then \rho,\, \sigma,\, \gamma \rangle \ltrans{\alpha} \langle \rho,\, \sigma',\, \gamma \rangle}
\end{mathpar}
\end{table}

Table~\ref{tab:event-rules} gives the rules.
The $\mathit{effect}$ function computes the store update: \textsc{Evt-Assign} evaluates and writes to $\sigma$; \textsc{Evt-Call} passes the slot addresses $\bar{x}$ to $g$ by reference, $g$ being opaque to the calculus and updating only those slots; \textsc{Evt-Seq} threads stores left-to-right through a sequence of statements.
The big-step rule \textsc{Evt} is the transitive closure of the statement-level steps, firing the entire event atomically as a single $\alpha$-labelled transition and hiding all intermediate stores from the environment.

\emph{In the diode.}
The diode computes nothing: forwarding a packet is a send.
The drivers do: \textsf{eth\_inner} hands each packet it receives on $\mathit{d2i}$ to its network card through an event whose block calls the C function the ethernet example already has (Figure~\ref{fig:diode-side}, line~\ref{c:send-frame}), which copies the frame into the transmit ring, cleans the cache, fills the descriptor and kicks the device, computation of a kind the calculus leaves to C:
\begin{lstlisting}[numbers=none]
import "eth.h" as eth
evt tx(x) = eth.send_frame(x)
\end{lstlisting}
The diode has no guard; the drivers' guards, whether a frame was polled and what it is, are tests over the local store their polling event fills.

\subsection{Processes}\label{sec:processes}
A process sequences events; a channel event is written the same way on both kinds of channel, the declared channel $c$ determining whether the communication is synchronous or asynchronous.

\noindent
\par\vspace{0.3em}\begin{bnf}
$\mathit{evt}$ ::= $\alpha$ : computation event (a named statement block)
             | $c \;?\; x$ : channel receive, binding the received value to variable $x$
             | $c \;!\; e$ : channel send, writing expression $e$ to channel $c$
\end{bnf}\smallskip

\noindent
\par\vspace{0.3em}\begin{bnf}
$Process$ $\rho$ ::= $\Omega$ : divergence ($\triangleq \mu X.\, X$)
                   | $Skip$ : successful termination of a process
                   | $\mathit{evt} \then \rho$ : event prefixing
                   | $[b]$ $\rho$ : guarded process
                   | $\rho_1$ $\extchoice$ $\rho_2$ : external choice
                   | $(\rho_1 \extchoice \rho_2) \;\mathbf{default}\; \rho_d$ : default external choice
                   | $\rho_1$ ; $\rho_2$ : sequential composition
                   | $\mu$ $X.$ \: $\rho$ : recursion with process variable $X$
\end{bnf}\smallskip

Divergence $\Omega \triangleq \mu X.\, X$ is the embedded failure mode: a component hangs on error rather than terminate, keeping shared state intact.
$\mathit{Skip}$ is a \emph{terminal} term: it performs no event, and termination is a predicate on states, defined below.
Recursion is otherwise required to be \emph{guarded}: in $\mu X.\,\rho$, every occurrence of $X$ in $\rho$ lies under an event prefix, so each unfolding performs at least one event before recurring; a prefix on the left of $\rho_1\,;\,\rho_2$ does not guard $\rho_2$, so a loop is written prefix-headed, $\mu X.\,\mathit{evt} \then (\cdots\,;\,X)$. Recursion is moreover \emph{tail} recursion: a recursion variable occurs only in tail position and refers to its own innermost loop. $\Omega$ is the sole unguarded exception, with its own dedicated rule. The variable $X$ is not surface syntax: it occurs only bound by $\mu X$. Named process definitions and calls, as in the running example, are surface sugar for recursion.
The default external choice $(\rho_1 \extchoice \rho_2)\;\mathbf{default}\;\rho_d$ guarantees progress: $\rho_d$ is taken when neither $\rho_1$ nor $\rho_2$ is enabled, ensuring the process always progresses.
When $\rho_1 \extchoice \rho_2$ is a sequence of guarded processes of the form $[b_1]\rho_1 \extchoice \cdots \extchoice [b_n]\rho_n$, the implicit guard of $\rho_d$ is $\neg(b_1 \vee \cdots \vee b_n)$, making the full partition complete by the law of excluded middle.
In particular, $(\rho_1 \extchoice \rho_2)\;\mathbf{default}\;Skip$ ends the current run of the body when nothing is enabled---the idiomatic ``yield'' of an event-driven component.
A choice of one branch, $[b]\,\rho \;\mathbf{default}\; \rho_d$, abbreviates $([b]\,\rho \extchoice [\bot]\,\rho)\;\mathbf{default}\;\rho_d$, with $\bot$ the guard that never holds.

\paragraph{Bodies}
A component program declares one body $\rho_{\mathit{body}}$, a process term whose run either terminates or diverges, $\Omega$ spinning on the event $\mathit{hang}$ (\textsc{Div}) and never returning to the platform; the component's initial local store is given, produced by whatever initialisation the platform runs before any communication.
How often a terminating body runs, and what runs it, is the platform's decision, not the term's: the process language has no top-level loop, and inner recursion is ordinary guarded $\mu X.\,\rho$ and may exit freely.
A body is \emph{receive-headed}: its head, through the left operand of $\,;\,$ and through $\mu$, is an external choice of receive prefixes, one per channel the component reacts to.
This is what makes a component reactive: a body runs because a message or an interrupt arrived, and its first act is to consume it; a polling loop is receive-headed on its device channel, a time-triggered task on its timer channel, and a component with nothing to wait on is not a component of this kind.
Receive-headedness keeps a quiet component always waiting on some channel, never silently quiescent, which the deadlock-freedom argument turns on.

\paragraph{Concurrency} Compared to classic CSP, we remove the parallel ($\rho_1 \parallel_{\alpha} \rho_2$) and interleave ($\rho_1 \interleave \rho_2$) compositions. We assume no concurrency within a component, consistent with the restriction of the platform we target; this remains a natural future extension.

Since the alphabet has no termination event, successful completion is a property of a local state, defined by structural descent on the process term:

\begin{definition}[Termination]\label{def:terminated}
The predicate $\terminated(\rho, \sigma, \gamma)$ holds exactly in the following cases:
\[
\begin{array}{l@{\quad\text{iff}\quad}l}
\terminated(\mathit{Skip}, \sigma, \gamma) & \text{always;} \\
\terminated([b]\,\rho, \sigma, \gamma) & \llbracket b \rrbracket_{\sigma} = \mathit{true} \;\wedge\; \terminated(\rho, \sigma, \gamma); \\
\terminated(\rho_1 \mathbin{;} \rho_2, \sigma, \gamma) & \terminated(\rho_1, \sigma, \gamma) \;\wedge\; \terminated(\rho_2, \sigma, \gamma); \\
\terminated((\rho_1 \extchoice \rho_2) \mathbin{\mathbf{default}} \rho_d, \sigma, \gamma) & \langle \rho_1, \sigma, \gamma \rangle \not\ltrans{\,} \;\wedge\; \langle \rho_2, \sigma, \gamma \rangle \not\ltrans{\,} \;\wedge\; \terminated(\rho_d, \sigma, \gamma); \\
\terminated(\mu X.\, \rho, \sigma, \gamma) & \terminated(\rho, \sigma, \gamma);
\end{array}
\]
and is false for all other process forms ($\Omega$, event prefixes, plain external choice, and $X$).
\end{definition}

A terminated state is quiet by \emph{success}: termination and enabledness are disjoint (a terminated process has no transitions), so \textsc{Seq} and \textsc{Seq-Term} below never overlap.
A $\mathbf{default}\,\mathit{Skip}$ choice is terminated precisely when neither branch is enabled, the ``nothing to do'' exit of a run of the body, after which the component is parked (\textsc{MK-Restart}); plain external choice never terminates.
The negative premises here and in \textsc{Ext-Default} are decided by the same structural descent, which guardedness of recursion keeps well-founded; the recursion clause descends into the body without unfolding.
Concretely, $\langle \rho, \sigma, \gamma \rangle \not\ltrans{}$ abbreviates that $\rho$ is not \emph{enabled}, where enabledness is defined by structural descent alongside $\terminated$: $\mathit{Skip}$ and a bound variable are not enabled, $\Omega$ is; a computation prefix is always enabled, a send when its channel is declared asynchronous (\textsc{Send} or \textsc{Loss}), a receive when its buffer is non-empty; $[b]\,\rho$ when $b$ holds and $\rho$ is enabled; a choice, default or not, when a branch is, a default choice also when $\rho_d$ is; $\rho_1\,;\,\rho_2$ when $\rho_1$ is enabled, or $\rho_1$ is terminated and $\rho_2$ is enabled; and $\mu X.\,\rho$ when $\rho$ is.
The $\rho_1$-terminated clause is the mutual recursion between enabledness and termination that the stratification orders.

\begin{table}[h]
\scriptsize
\caption{Process rules}\label{tab:process-rules}
\begin{mathpar}
\inferrule*[left=\textsc{Div}]
{\mathit{hang} \in \Sigma_\alpha}
{\langle \Omega,\, \sigma,\, \gamma \rangle \ltrans{\mathit{hang}} \langle \Omega,\, \sigma,\, \gamma \rangle}

\inferrule*[left=\textsc{Guard-T}]
{\llbracket b \rrbracket_{\sigma} = \mathit{true} \\
  \langle \rho,\, \sigma,\, \gamma \rangle \ltrans{\ell} \langle \rho',\, \sigma',\, \gamma' \rangle}
{\langle [b]\, \rho,\, \sigma,\, \gamma \rangle \ltrans{[b]\cdot\ell} \langle \rho',\, \sigma',\, \gamma' \rangle}

\inferrule*[left=\textsc{ExtL}]
{\langle \rho_1,\, \sigma,\, \gamma \rangle \ltrans{\ell} \langle \rho_1',\, \sigma',\, \gamma' \rangle}
{\langle \rho_1 \extchoice \rho_2,\, \sigma,\, \gamma \rangle \ltrans{\ell} \langle \rho_1',\, \sigma',\, \gamma' \rangle}

\inferrule*[left=\textsc{ExtR}]
{\langle \rho_2,\, \sigma,\, \gamma \rangle \ltrans{\ell} \langle \rho_2',\, \sigma',\, \gamma' \rangle}
{\langle \rho_1 \extchoice \rho_2,\, \sigma,\, \gamma \rangle \ltrans{\ell} \langle \rho_2',\, \sigma',\, \gamma' \rangle}

\inferrule*[left=\textsc{ExtDefault-L}]
{\langle \rho_1,\, \sigma,\, \gamma \rangle \ltrans{\ell} \langle \rho_1',\, \sigma',\, \gamma' \rangle}
{\langle \chi,\, \sigma,\, \gamma \rangle \ltrans{\ell} \langle \rho_1',\, \sigma',\, \gamma' \rangle}

\inferrule*[left=\textsc{ExtDefault-R}]
{\langle \rho_2,\, \sigma,\, \gamma \rangle \ltrans{\ell} \langle \rho_2',\, \sigma',\, \gamma' \rangle}
{\langle \chi,\, \sigma,\, \gamma \rangle \ltrans{\ell} \langle \rho_2',\, \sigma',\, \gamma' \rangle}

\inferrule*[left=\textsc{Ext-Default}]
{\langle \rho_1,\, \sigma,\, \gamma \rangle \not\ltrans{\,} \qquad
  \langle \rho_2,\, \sigma,\, \gamma \rangle \not\ltrans{\,} \\
  \langle \rho_d,\, \sigma,\, \gamma \rangle \ltrans{\ell} \langle \rho_d',\, \sigma',\, \gamma' \rangle}
{\langle \chi,\, \sigma,\, \gamma \rangle \ltrans{\ell} \langle \rho_d',\, \sigma',\, \gamma' \rangle}

\inferrule*[left=\textsc{Seq}]
{\langle \rho_1,\, \sigma,\, \gamma \rangle \ltrans{\ell} \langle \rho_1',\, \sigma',\, \gamma' \rangle}
{\langle \rho_1 \mathbin{;} \rho_2,\, \sigma,\, \gamma \rangle \ltrans{\ell} \langle \rho_1' \mathbin{;} \rho_2,\, \sigma',\, \gamma' \rangle}

\inferrule*[left=\textsc{Seq-Term}]
{\terminated(\rho_1, \sigma, \gamma) \\
  \langle \rho_2,\, \sigma,\, \gamma \rangle \ltrans{\ell} \langle \rho_2',\, \sigma',\, \gamma' \rangle}
{\langle \rho_1 \mathbin{;} \rho_2,\, \sigma,\, \gamma \rangle \ltrans{\ell} \langle \rho_2',\, \sigma',\, \gamma' \rangle}

\inferrule*[left=\textsc{Rec}]
{\langle \rho[\mu X.\: \rho / X],\, \sigma,\, \gamma \rangle \ltrans{\ell} \langle \rho',\, \sigma',\, \gamma' \rangle}
{\langle \mu X.\: \rho,\, \sigma,\, \gamma \rangle \ltrans{\ell} \langle \rho',\, \sigma',\, \gamma' \rangle}
\end{mathpar}
\end{table}

\paragraph{Rules.}
Table~\ref{tab:process-rules} gives the rules, writing $\chi$ for the default choice $(\rho_1 \extchoice \rho_2)\;\mathbf{default}\;\rho_d$.
Structural reductions---sequential termination, recursion unfolding, guard evaluation and default fallthrough---are each combined with the first event of the continuation into a single step (\textsc{Seq-Term}, \textsc{Rec}, \textsc{Guard-T}, \textsc{Ext-Default}), eliminating the intermediate $\tau$-transitions of standard CCS/CSP presentations; $\mathit{Skip}$ has no rule, and $\Omega$ spins on $\mathit{hang}$ (\textsc{Div}).
The atomicity of \textsc{Guard-T} is essential, not cosmetic.
The classical presentation emits a $\tau$-step once the guard evaluates to true, leaving the body enabled; in a choice between guarded branches that $\tau$-step resolves the choice, silently and before any event is observed, so two branches whose guards both hold make the choice internal.
Evaluating the guard and firing the event in one step keeps the resolution on the visible label, which is what the determinism of \S\ref{sec:proofs}, at most one successor per label, and the absence of $\tau$ rest on; nested guards, as arise when a guarded branch has a guarded body, flatten by the guard merge, so every label of $\Lambda^p$ bears at most one guard.
External choice is resolved by the first event of either branch (\textsc{ExtL}/\textsc{ExtR}), with no $\ell \neq \tau$ side condition since $\tau \notin \Sigma^p$; a default choice resolves likewise (\textsc{ExtDefault-L}/\textsc{R}), and when neither branch is enabled \textsc{Ext-Default} fires the first event of $\rho_d$, the structural dual of the priority operator of Cleaveland and Hennessy~\cite{cleaveland90priority}.
Its negative premises are well-defined under stratified SOS~\cite{bol96negative}: stuckness is decided by the structural descent above, on strictly smaller terms, and guardedness forbids a recursion variable at a branch head, so no derivation cycles through the rule's own negation.
\textsc{Seq} runs the left process, and \textsc{Seq-Term} fires the first event of the right once the left is terminated, a property of the state rather than a syntactic match on $\mathit{Skip}$; \textsc{Rec} unfolds and fires the first event of the unfolding in one step.

\emph{In the diode.}
The diode's work on one packet is a receive followed by a send, $\mathit{o2d}?x \then \mathit{d2i}!x \then \mathit{Skip}$.
The receive takes the oldest value off $\mathit{o2d}$ and blocks while there is none; the send appends $x$ to $\mathit{d2i}$ and, when $\mathit{d2i}$ is full, drops it and carries on (\textsc{Send} and \textsc{Loss}).
The C's full test, its drop branch, its copy and its flag updates (lines~\ref{c:full}--\ref{c:send-to} and~\ref{c:release}) are exactly these two events.
That is the whole component program, the C's \lstinline|init| (line~\ref{c:init}) having nothing to initialise:
\begin{lstlisting}[numbers=none]
component dd
body = o2d?x -> d2i!x -> Skip
\end{lstlisting}
The C's \lstinline|for| loop, which moves every packet present when the handler is called, has no counterpart, and needs none: a body that has terminated is run again whenever a packet is there to receive, which is the platform's rule rather than the program's (\S\ref{sec:platform}), so one packet per run drains the ring; the C loops because one notification may stand for several packets, a fact of the platform that the program should not have to know.

The body starts with a receive where the C starts with \lstinline|switch (ch)| (line~\ref{c:switch}): a component with several input channels starts with a choice of receives, one per channel.
The inner driver reacts to two channels, the frames on $\mathit{d2i}$ and its network card's interrupt; it transmits a frame through the event $\mathit{tx}$ of \S\ref{sec:comp-events}, and serves an interrupt before acknowledging it to the device:
\begin{lstlisting}[numbers=none]
component eth_inner
body = d2i?x -> tx(x) -> Skip
    [] irq?n -> handle_irq -> irq!n -> Skip
\end{lstlisting}
The C's outer \lstinline|for| and \lstinline|break| are gone, and so are the indices, the flags and the addresses; what is left is the diode's one decision, forward or drop, and the drop is the channel's.
$\Omega$ has a C counterpart in the same example: on a zero-length packet, the ethernet driver stops in \lstinline|for (;;) { }|.

\subsection{Local Labelled Transition System}\label{sec:lts}
The rules above define, for each component, a transition system over its local states; a component is a self-contained process that operates independently unless communication arises, channel events being left as blocking points that the composition resolves.

\begin{definition}[Local Labelled Transition System]\label{def:local-automaton}
The \emph{local labelled transition system} ($llts$) of component $p$ is the tuple
$(S^p,\; \Sigma^p,\; {\to},\; s^p_0)$ where:
\begin{itemize}
\item $S^p$ is the set of local states $\langle \rho^p,\, \sigma^p,\, \gamma^p \rangle$;
\item $\Sigma^p$ is the event alphabet, which contains \emph{no unobservable event}: neither $\tau$ nor a termination marker $\checkmark$;
\item ${\to} \subseteq S^p \times \Lambda^p \times S^p$ is the transition relation defined by the rules of Tables~\ref{tab:comm-rules}--\ref{tab:process-rules}, where $\Lambda^p = \Sigma^p \cup \{\, [b]\,\ell \mid \ell \in \Sigma^p,\ b \text{ a boolean guard} \,\}$ closes the alphabet under guard annotation; and
\item $s^p_0 = \langle \mathit{Skip},\, \sigma^p_0,\, \gamma^p_0 \rangle$ is the initial state: the component is parked between runs of its body, its residual terminated, and its first step is the first step of $\rho^p_{\mathit{body}}$ under the platform's rule (\textsc{MK-Restart}); $\sigma^p_0$ is the component's initial store and $\gamma^p_0 = \Gamma_0|_{C_p}$ its view of the initial buffers, both given with the initial configuration.
\end{itemize}
\end{definition}

\begin{figure*}[t]
\centering
\begin{tikzpicture}[scale=0.62, transform shape,
  >=Stealth,
  every state/.style={minimum size=8mm, inner sep=1pt, font=\scriptsize},
  ev/.style={font=\scriptsize, text=cyan, align=center},
  stlbl/.style={font=\scriptsize, text=black}]

\node[font=\small\bfseries] at (3.1, 1.9) {\textsf{dd}};
\node[state] (d0) at (0,   0) {$s_0$};
\node[state] (d1) at (6.2, 0) {$s_1$};
\draw[->] (-0.6,0.6) -- (d0);
\path[->]
  (d0) edge[bend left=20]  node[ev,above]{$o2d?x$}  (d1)
  (d1) edge[bend left=20]  node[ev,below]{$d2i!x$}  (d0);
\node[stlbl,above=3pt of d0]{$\rho_{\mathit{body}}$};
\node[stlbl,above=3pt of d1]{Forward};

\begin{scope}[xshift=9cm]
\node[font=\small\bfseries] at (3.1, 1.6) {\textsf{eth\_inner}};
\node[state] (i0) at (3.1, -1.2) {$s_0$};
\node[state] (i1) at (0,    0) {$s_1$};
\node[state] (i2) at (6.2,  0) {$s_2$};
\node[state] (i3) at (0,   -2.4) {$s_3$};
\path[->]
  (i0) edge                node[ev,below,sloped,pos=0.5]{$\mathit{irq?78}$}      (i1)
  (i1) edge                node[ev,above]{$\mathit{handle\_irq}$}                (i2)
  (i2) edge                node[ev,below,sloped,pos=0.5]{$\mathit{irq!78}$}      (i0)
  (i0) edge                node[ev,above,sloped,pos=0.5]{$d2i?x$}                (i3)
  (i3) edge[bend right=20] node[ev,below,sloped,pos=0.5]{$\mathit{tx}(x)$}     (i0);
\draw[->] (3.1,-0.3) -- (i0);
\node[stlbl,below=3pt of i0]{Notified};
\node[stlbl,above=1pt of i1]{Handle\_eth};
\node[stlbl,above=1pt of i2]{Rearm};
\node[stlbl,below=1pt of i3]{Deliver};
\end{scope}

\begin{scope}[yshift=-5.8cm]
\node[font=\small\bfseries] at (0.8, 1.0) {\textsf{eth\_outer}};
\node[state]          (e1) at (3,   0) {$s_1$};
\node[state]          (e2) at (7.5, 0) {$s_2$};
\node[state, initial left] (e0) at (3, -2.6) {$s_0$};
\node[state]          (e3) at (7.5, -2.6) {$s_3$};
\node[state]          (e5) at (13, -2.6) {$s_5$};
\path[->]
  (e0) edge                 node[ev,left]{$\mathit{irq?79}$}                                                   (e1)
  (e1) edge                 node[ev,above]{$\mathit{handle\_irq}$}                                             (e2)
  (e2) edge                 node[ev,left]{$\mathit{poll}$}                                                     (e3)
  (e3) edge[bend right=40]  node[ev,right,pos=0.5]{$\textcolor{guardgreen}{[\mathit{rx} \wedge \mathit{pass}]}$\\$o2d!\mathit{pkt}$} (e2)
  (e3) edge                 node[ev,below,pos=0.45]{$\textcolor{guardgreen}{[\mathit{rx} \wedge \neg\mathit{pass} \wedge \mathit{reply}]}\ \mathit{handle\_ping}$} (e5)
  (e5) edge                 node[ev,right,pos=0.85]{$\mathit{tx}(\mathit{reply})$}               (e2)
  (e3) edge[loop below]     node[ev]{$\textcolor{guardgreen}{[\mathit{rx} \wedge \neg\mathit{pass} \wedge \neg\mathit{reply}]}\ \mathit{poll}$} (e3)
  (e3) edge                 node[ev,below]{$\textcolor{guardgreen}{[\neg\mathit{rx}]}\ \mathit{irq!79}$}       (e0);
\node[stlbl,below=3pt of e0]{Notified};
\node[stlbl,above=1pt of e1]{Handle\_eth};
\node[stlbl,above=1pt of e2]{Next\_pkt};
\node[stlbl,above left=1pt of e3]{Handle\_rx};
\node[stlbl,below=3pt of e5]{Reply};
\end{scope}

\end{tikzpicture}
\caption{STGs of the diode components.}
\label{fig:llts-diode}
\end{figure*}
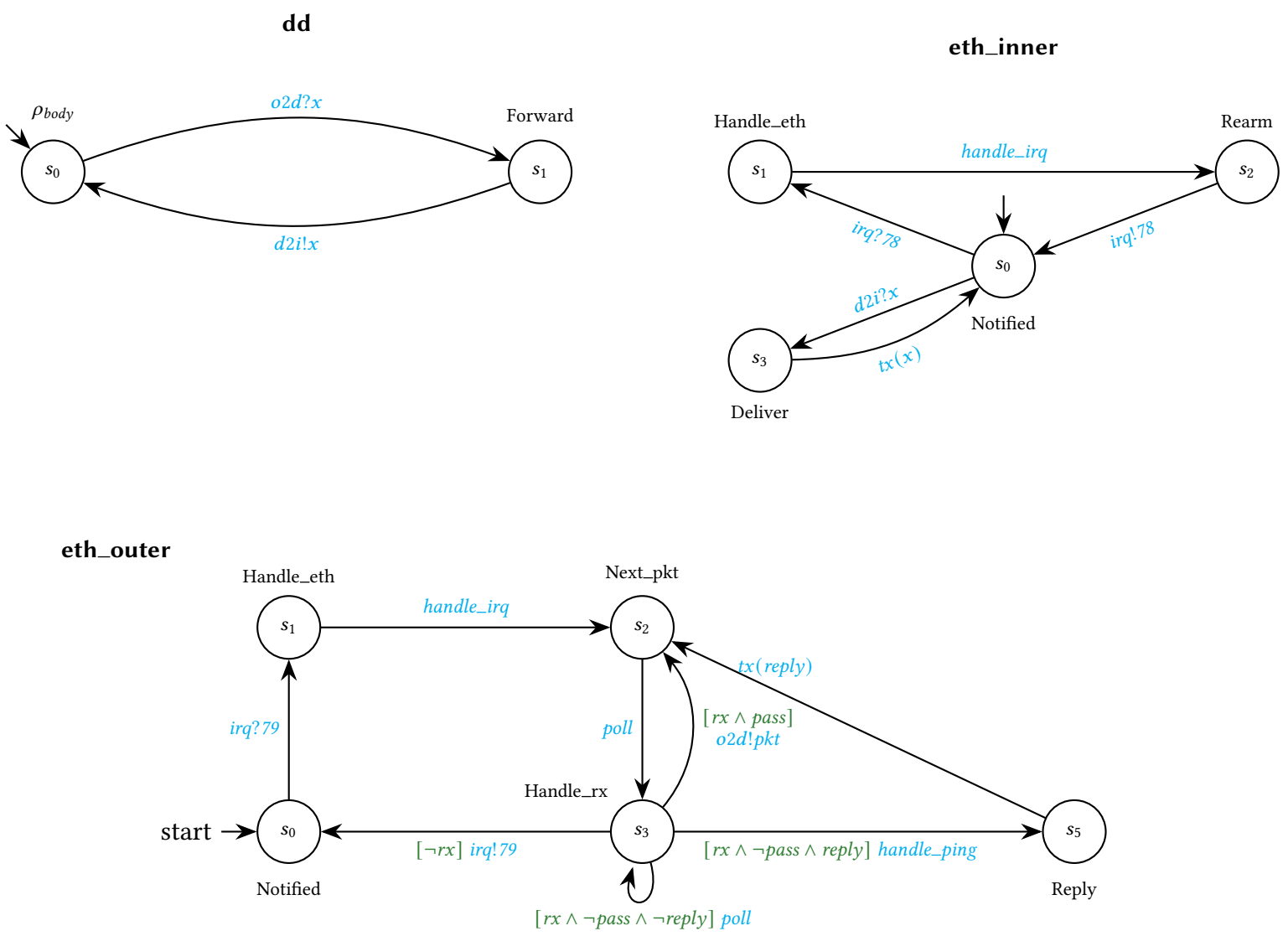

\paragraph{Example: Data Diode Application.}
Figure~\ref{fig:llts-diode} draws each component as a \emph{symbolic transition graph} (STG)~\cite{hennessy95symbolic}---a finite graph whose nodes are process terms and whose edges carry (possibly guarded) events---derived from the component programs of the running example (the diode's is the program assembled above).
The STG is the finite, syntax-level picture; interpreting it over the data state generates the concrete \emph{llts}, the object our results are stated on, and all LTS figures in this paper follow this convention; the term-to-STG construction is standard for value-passing calculi~\cite{hennessy95symbolic,lin96stga} and not a contribution here.
\textsf{dd} receives on $o2d$ and forwards on $d2i$, the send dropping when the downstream buffer is full; the two ethernet components serve interrupts and their device rings in the same receive-then-act shape.
State names are illustrative; the guards of the ethernet components read $\mathit{rx}$, a frame was polled, $\mathit{pass}$, foreign traffic to forward, and $\mathit{reply}$, ARP or ICMP addressed to this host, and the diode has none.

\section{System Semantics}\label{sec:system}

\subsection{Global Composition}\label{sec:global-comp}
A system is the interleaving of its component processes, each described by its \emph{llts}, over the shared channel buffers: a component's computation events and asynchronous channel events are lifted to the global level as they are, and a rendezvous is \textsc{Sync} applied to two components.
The interleaving is parameterised on the platform in two places, \emph{which} process a component offers next, its \emph{invocation} $\mathit{inv}(s^p)$, and the configuration the system starts from, $\mathcal{G}_0$; the next subsection supplies both for seL4 MicroKit.
Per-component alphabets are pairwise disjoint, an event name being qualified by its component, so a lifted label identifies the component that fired it.

\begin{definition}[Global Configuration]\label{def:global-config}
A \emph{global configuration} is a pair $\langle \mathbf{S},\, \Gamma \rangle$ with $\mathbf{S} = (s^p)_{p \in P}$ one local state per component and $\Gamma$ the global channel store, the view of every component read off the store, \eqref{inv:gcoh}; the initial configuration $\mathcal{G}_0 = \langle \mathbf{S}_0,\, \Gamma_0 \rangle$ is given with the system: every component parked, $s^p_0 = \langle \mathit{Skip}, \sigma^p_0, \Gamma_0|_{C_p} \rangle$, over initial stores and buffers within capacity.
\end{definition}

The invocation $\mathit{inv}(s^p)$ is the residual while it has not terminated and a fresh body once it has (\textsc{MK-Restart} below), so running the body again is not a transition of its own but is folded into the component's next step, exactly as \textsc{Seq-Term} and \textsc{Rec} fold a structural reduction into the next event within a term, and a parked component offers what its fresh body offers.

\begin{definition}[Global LTS]\label{def:global-lts}
The \emph{glts} $\zeta \triangleq \Interleave_{p \in P}\, \mathit{llts}_p$ is the LTS $(\mathcal{G},\, \Sigma_g,\, \to_g,\, \mathcal{G}_0)$ where $\mathcal{G}$ is the set of global configurations satisfying~\eqref{inv:gcoh};
the global alphabet $\Sigma_g = \bigcup_{p \in P} \Sigma^p \;\cup\; \{\, c.v \mid c \in C_{\mathit{sync}},\, v \in \mathit{Val} \,\}$ comprises the local events of every component together with the rendezvous labels $c.v$ of synchronous channels; and ${\to_g}$ is the least relation closed under the rules of Table~\ref{tab:sos-compose}.
\end{definition}

\begin{table}[t]
\scriptsize
\centering
\begin{mathpar}
\inferrule*[left=\textsc{Ileave}]
{\mathit{inv}(s^p) \ltrans{\lambda} {s^p}' \quad \lambda \in \{\, \ell,\ [b]\,\ell \,\} \quad \ell \in \Sigma^p}
{\langle \mathbf{S},\, \Gamma \rangle \ltrans{\ell}
  \langle \mathbf{S}[p \mapsto {s^p}'],\; \Gamma \oplus \gamma^{p\prime} \rangle}
\\
\inferrule*[left=\textsc{G-Sync}]
{\mathit{inv}(s^{p_1}) = \langle [b_1]\,\rho_1,\, \sigma_1,\, \gamma_1 \rangle \quad
  \mathit{inv}(s^{p_2}) = \langle [b_2]\,\rho_2,\, \sigma_2,\, \gamma_2 \rangle \\
  \llbracket b_1 \rrbracket_{\sigma_1} = \mathit{true} \quad
  \llbracket b_2 \rrbracket_{\sigma_2} = \mathit{true} \\
  (\langle \rho_1,\, \sigma_1,\, \gamma_1 \rangle,\; \langle \rho_2,\, \sigma_2,\, \gamma_2 \rangle) \ltrans{c.v} (s_1',\, s_2') \quad (\textsc{Sync})}
{\langle \mathbf{S},\, \Gamma \rangle \ltrans{c.v}
  \langle \mathbf{S}[p_1 \mapsto s_1',\; p_2 \mapsto s_2'],\; \Gamma \rangle}
\end{mathpar}
\caption{Global transition rules.}
\label{tab:sos-compose}
\end{table}

\textsc{Ileave} lifts any local step of the invocation of a component $p$---a computation event, an asynchronous send (\textsc{Send} or \textsc{Loss}) or receive---to the global level; $\Gamma \oplus \gamma^{p\prime}$ is $\Gamma$ updated on $C_p$ to the new buffer view of $p$, so that \eqref{inv:gcoh} is maintained by construction.
A guard is checked at the moment its event fires and is not an observable of its own, so \textsc{Ileave} lifts $[b]\,\ell$ as $\ell$: the traces of $\zeta$ are sequences over $\Sigma_g$ alone.
\textsc{G-Sync} applies \textsc{Sync} to the invocations of the two endpoints when both guards hold (unguarded case: $b_1 = b_2 = \top$); it is shown for the head-normal case, and in general the two offers are matched at the enabled frontiers of the two invocations.
Computation events remain labels of $\zeta$ rather than being hidden behind $\tau$: no other component synchronises on them, the results of \S\ref{sec:proofs} are stated on $\zeta$ as defined, and keeping them visible keeps the composition at the granularity of the program; hiding them, the classical CSP route to a smaller state space~\cite{roscoe2010understanding}, is the abstraction of \S\ref{sec:pd-global}, a refinement beside this composition rather than a step of it.

\subsection{The Platform: seL4 MicroKit}\label{sec:platform}
We instantiate the calculus on the verified microkernel seL4~\cite{klein09sel4} and its MicroKit development framework~\cite{weibel24}.
MicroKit exports seL4's guarantees as four abstractions: the \emph{protection domain}, an isolated process; the \emph{channel}, wrapping inter-process communication and shared memory; the \emph{memory region}; and the \emph{mapping} of a region into a protection domain with access rights.
It provides only this static configuration, every behaviour being written in C, and leaves the mappings unrestricted: a developer can grant a third party access to a buffer meant for two communicating domains.

\paragraph{Memory regions and mappings.}
A memory region $m = (\mathit{size}, \mathit{pbase})$ declares a page-aligned byte size and a base physical address, user provided or allocated at boot.
Each component declares its mappings $\mathcal{M}^p$, each $\mu = (m,\, \mathit{vaddr},\, \omega,\, \mathit{slotsize})$ binding a region at a virtual base address with access rights $\omega \subseteq \{\mathit{r},\mathit{w},\mathit{x}\}$ and a slot granularity, as global variables of its events and processes; the mechanisation abstracts a mapping to the region it grants, the rest being the platform's to enforce.
The diode declares no mapping, touching only its two channel buffers, which the platform maps into both ends; a driver maps its network card's registers.
That mapping of a buffer region, with write access for the sender and read access for the receiver, is what \textsc{Chan-Async} leaves to the platform:
\begin{mathpar}\scriptsize
\inferrule*[left=\textsc{Chan-Map}]
{c \triangleq (p_1 \to_m^{N} p_2) \in C \qquad
  \mu_1 = (m,\, \mathit{vaddr}_1,\, \{\mathit{r},\mathit{w}\}) \qquad
  \mu_2 = (m,\, \mathit{vaddr}_2,\, \{\mathit{r}\})}
{\mathcal{M}^{p_1},\; \mathcal{M}^{p_2} \vdash c \leadsto \textsc{Map}(\mu_1,\, p_1),\; \textsc{Map}(\mu_2,\, p_2)}
\end{mathpar}
In the running example, neither ethernet component maps the other's channel buffer.

\begin{figure}[t]
\centering
\begin{tikzpicture}[font=\scriptsize, >=Stealth]

\begin{scope}[shift={(0,1.7)}]
  \node[anchor=south west, font=\scriptsize\bfseries] at (-0.05, 0.55)
    {component $p_1$ \; ($\mu^{p_1}$, \ $\omega_1 = \{r,w\}$)};
  \foreach \i in {0,...,4} {
    \draw (\i*0.7, 0) rectangle ++(0.7, 0.5);
  }
  \draw[fill=pdblue!20] (1.4, 0) rectangle ++(0.7, 0.5);
  \node at (1.75, 0.25) {$i$};
  \node[anchor=north west, inner sep=1pt] at (-0.05, -0.07) {$\mathit{vbase}_1$};
\end{scope}

\begin{scope}[shift={(5.2,1.7)}]
  \node[anchor=south west, font=\scriptsize\bfseries] at (-0.05, 0.55)
    {component $p_2$ \; ($\mu^{p_2}$, \ $\omega_2 = \{r\}$)};
  \foreach \i in {0,...,4} {
    \draw (\i*0.7, 0) rectangle ++(0.7, 0.5);
  }
  \draw[fill=pdblue!20] (1.4, 0) rectangle ++(0.7, 0.5);
  \node at (1.75, 0.25) {$i$};
  \node[anchor=north west, inner sep=1pt] at (-0.05, -0.07) {$\mathit{vbase}_2$};
\end{scope}

\begin{scope}[shift={(2.6,0)}]
  \foreach \i in {0,...,4} {
    \draw (\i*0.7, 0) rectangle ++(0.7, 0.5);
  }
  \draw[fill=floworange!25] (1.4, 0) rectangle ++(0.7, 0.5);
  \node at (1.75, 0.25) {$i$};
  \node[anchor=north west, inner sep=1pt] at (-0.05, -0.07) {$\mathit{pbase}(m)$};
  \node[anchor=west] at (3.6, 0.25)
    {region $m$ \ (slots of size $\mathit{slotsize}(m)$)};
  \node[anchor=north, inner sep=2pt] at (1.75, -0.35)
    {$\mathit{paddr}(m, i)$};
\end{scope}

\draw[->, dashed] (1.75, 1.7) -- (4.0, 0.55)
  node[midway, left=2pt] {$\mathit{MMU}_{p_1}$};
\draw[->, dashed] (6.95, 1.7) -- (4.7, 0.55)
  node[midway, right=2pt] {$\mathit{MMU}_{p_2}$};

\node[anchor=north] at (4.35, -0.75)
  {$\mu^{p_1}(i) \;=\; \mu^{p_2}(i) \;=\; m(i)$ \quad \eqref{inv:coh}};

\end{tikzpicture}
\caption{The memory model.}
\label{fig:memory-model}
\end{figure}
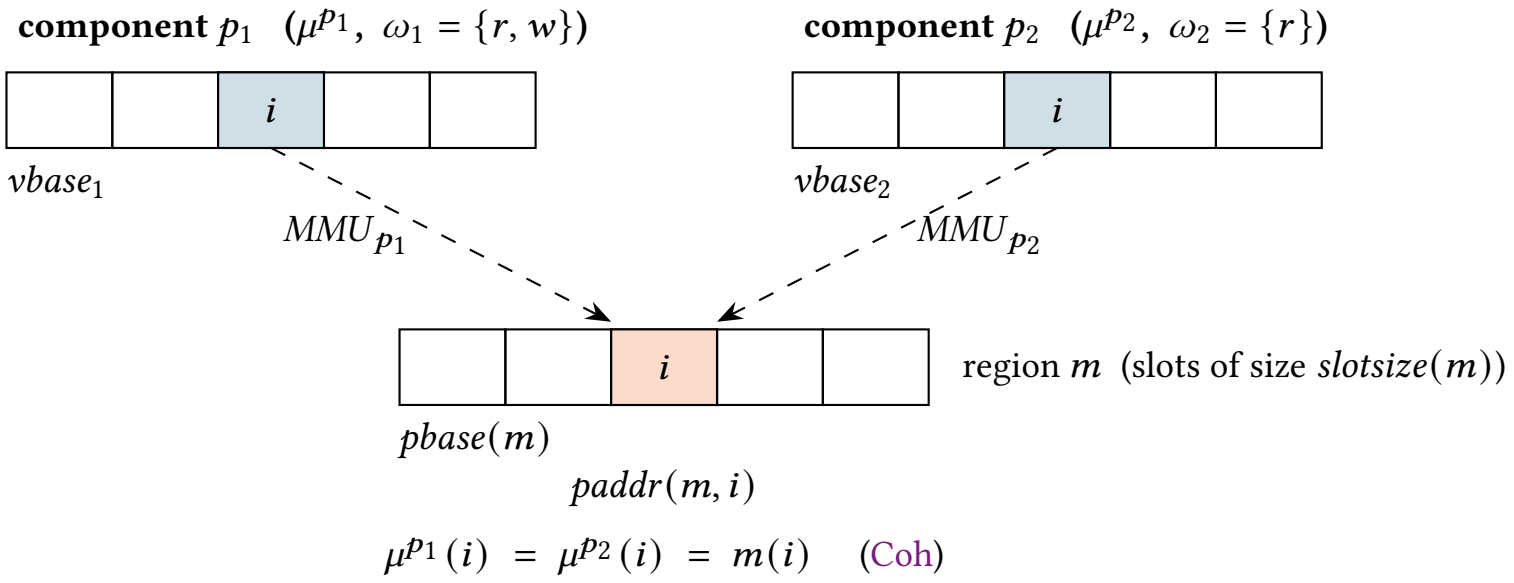

\paragraph{Memory model.}\label{sec:mappings}
Memory is a set $M$ of \emph{regions}, each a partial function $m : \mathbb{N} \rightharpoonup \mathit{Val}$ from slot index to value, slot $i$ of $m$ having the physical address $\mathit{paddr}(m, i)$.
A component $p$ maps the regions it may access, at most one mapping per region; they are its \emph{accessible} regions $M^p$, split into its private regions and the buffers $M^C$ of the channels it is an endpoint of.
The memory-management unit translates every mapping of a region to the same physical slots (Figure~\ref{fig:memory-model}), and two invariants follow, which are all the semantics uses: the mappings of one component do not overlap, and every endpoint of a channel sees the same buffer,
\begin{gather}
\forall\, \mu \neq \mu' \in \mathcal{M}^p.\;\; \mathit{range}(\mu) \cap \mathit{range}(\mu') = \emptyset
\tag{Disj}\label{inv:disj}
\\
\mu^{p_1}(i) \;=\; \mu^{p_2}(i) \;=\; m(i)
\tag{Coh}\label{inv:coh}
\end{gather}
where $\mu^{p_1}$ and $\mu^{p_2}$ map the same region $m$.

\paragraph{Protection domains as components.}
A protection domain is a component of the calculus: an address space, the component's private store with the regions it maps; a priority $p.\mathit{pri}$, distinct from every other's; and its channels and mappings.
Mappings and rights are fixed at boot and enforced by seL4's capabilities; CoreFi declares the grants and never checks them.
A hardware interrupt is an interrupt channel between the device and its protection domain, delivered as a notification and acknowledged by \lstinline|microkit_irq_ack|.
A synchronous channel is MicroKit's protected procedure call, allowed only to a domain of strictly higher priority, the further constraint \textsc{Chan-Sync} leaves to the platform:
\begin{mathpar}\scriptsize
\inferrule*[left=\textsc{Chan-Call}]
{c \triangleq (p_1 \rightarrowtail p_2) \in C \qquad p_1.\mathit{pri} < p_2.\mathit{pri}}
{c \leadsto \mathit{ok}}
\end{mathpar}
Since priorities are totally ordered, \textsc{Chan-Call} alone excludes every cycle of synchronous calls; the cycles that mix channel kinds are (\textsc{Wf}1)'s to exclude.

\paragraph{Entry points.}
A protection domain has two entry points: \lstinline|init()|, run once before any notification, which produces the domain's initial store and is not modelled beyond that, and \lstinline|notified()|, which is the body $\rho_{\mathit{body}}$, each run of which is one invocation, re-run by the kernel on the next notification once it has terminated; a diverging body never returns.
The \lstinline|notified(channel)| contract, that an invocation exists only because a channel fired, is receive-headedness: a top-level guard would make the wake-up conditional and a top-level $\mathbf{default}$ would poll, where MicroKit parks an idle domain, so guards, computation events, sends and $\mathbf{default}$ escapes appear only after the initial receive.

\paragraph{The scheduler.}
A protection domain's process term is the residual of its current invocation of \lstinline|notified|; when that residual is terminated the domain has returned to the kernel, and its next step is the first step of a fresh run of its body:
\[
\mathit{inv}(s^p) \;\triangleq\;
\begin{cases}
\langle \rho^p_{\mathit{body}},\, \sigma^p,\, \gamma^p \rangle & \text{if } \terminated(\rho^p, \sigma^p, \gamma^p) \\
s^p & \text{otherwise.}
\end{cases}
\]
This is the rule \textsc{MK-Restart}.
The initial configuration $\mathcal{G}_0$ is what MicroKit's boot leaves: it runs the \lstinline|init()| of every domain in decreasing priority order from empty buffers, after which every domain is parked over the store its \lstinline|init()| produced; we take $\mathcal{G}_0$ as given, every component parked and the buffers within capacity, which such a boot guarantees.

\paragraph{The system composition.}
A MicroKit system is declared by a \emph{system description}, which lists the memory regions, the protection domains and the channels, and composes the protection domains by interleaving, $\zeta = \Interleave_{p \in P}\, p$: CSP's interleaving without handshaking, every interaction being through a declared channel, whose semantics is the global composition of \S\ref{sec:global-comp} with the invocation and the initial configuration just defined.

\subsection{Beyond MicroKit}\label{sec:generality}
Nothing in the global composition is MicroKit's but the two parameters it reads, the invocation $\mathit{inv}$ and the initial configuration $\mathcal{G}_0$; another platform is another pair.
The results of \S\ref{sec:proofs} use two facts about the pair, that an invocation is a fresh body or the residual of one, and that the initial configuration satisfies the buffer invariant; a platform whose pair has them inherits the theorems, and the mechanisation proves them for MicroKit's.
A framework with the same run-to-completion handlers but a different entry-point discipline changes only the contracts on its bodies; a cooperative scheduler that lets a body yield mid-run supplies an $\mathit{inv}$ that returns the residual at the yield rather than at termination; a time-triggered platform, as Mimosa's is~\cite{huber25mimosa}, supplies an invocation keyed to a clock, where a timing extension would start.
The invocation can also be written as a term over the component bodies with the composition operators CSP already has: MicroKit's steady state is the interleaving of the bodies under iteration, $\Interleave_{p \in P}\, {*}\rho^p_{\mathit{body}}$, where ${*}\rho$ restarts $\rho$ once it has terminated; a round-robin scheduler is a sequential cycle of the bodies instead, and a preemptive one a prioritised choice between them~\cite{cleaveland90priority}.
We keep the scheduler as a rule rather than a term because the rule is what the proofs read, and because a term would put the platform inside the components' process language, where a component could express a scheduling decision that is not its to make.

\section{Properties}\label{sec:proofs}

We prove three properties: every system keeps its channel buffers within capacity, and every well-formed system is deadlock-free and deterministic.
Deadlock freedom is a baseline requirement for any communication-based system---a stuck component can hang an entire embedded system indefinitely.
Determinism makes a run reproducible from its labels, which testing and certification rely on.
Both follow from a single structural discipline, \emph{well-formedness}.

\subsection{Well-formedness}\label{sec:well-formed}

Well-formedness is a set of structural conditions on the system.
Each channel has a designated sender and receiver, both components except on an interrupt channel, whose other end is the environment.
Beyond these declaration facts, two conditions---acyclicity and choice partition---must hold, and the definition also bundles the grammar and declaration facts on which the theorems rely.

\begin{definition}[Well-formed system]\label{def:well-formed}
For an external choice $\chi$ let $\mathit{br}(\chi) = \{(b_i,\ \ell_i \then \rho_i)\}_i$ be its \emph{flattened} branches, those of the choices nested in its branches included, nested guards merged by conjunction and $b_i = \top$ for unguarded branches; and write $\models\varphi$ when $\varphi$ holds in every store $\sigma$.
A system is \emph{well-formed} iff the following two conditions hold, together with the grammar and declaration facts listed below.
\begin{description}
  \item[(\textsc{Wf}1)] \emph{Acyclic dependency.}
  \begin{gather*}
    \forall p \in P.\;\; \neg\,(p \leadsto^{+} p), \qquad\text{where}\\
    p \leadsto q \;\iff\; \exists c \in C.\;
    \big(p = p_r(c) \wedge q = p_s(c)\big) \;\vee\;
    \big(c \in C_{\mathit{sync}} \wedge p = p_s(c) \wedge q = p_r(c)\big).
  \end{gather*}
  \item[(\textsc{Wf}2)] \emph{Choice partition.}
  Every guard is a branch of an external choice, every branch is event-headed, and for every choice $\chi$:
  \begin{itemize}
    \item \emph{branch disjointness}: $\forall i \ne j.\;\; \mathit{head}(\ell_i) = \mathit{head}(\ell_j) \;\Rightarrow\; {\models}\,\neg(b_i \wedge b_j)$;
    \item \emph{exhaustiveness}: if $\chi$ is a plain choice, ${\models}\, \bigvee_i b_i$;
    \item \emph{locally decidable defaults}: if $\chi$ is a $\mathbf{default}$ choice, every $\ell_i$ is a computation event or an asynchronous send, and $\rho_d$ is partitioned;
  \end{itemize}
  where $\mathit{head}(\ell)$ is the name of a computation event and the channel together with the direction of a channel event, values ignored, and $\Omega$ as a branch heads the computation event $\mathit{hang}$.
\end{description}
\end{definition}

The grammar and declaration facts bundled with the two conditions are: every body is closed, guarded in its recursion, and tail-recursive, with inner loops recurring only to themselves; every body is receive-headed and every channel other than an interrupt channel runs between two components; and every body sends only on channels it is the declared sender of and receives only on those it is the declared receiver of.

\paragraph{Remark: what each condition guards against.}
(\textsc{Wf}1) reads $p \leadsto q$ as ``$p$ may have to wait for $q$'': every receiver waits for its sender, and a synchronous sender additionally waits for its receiver, because a rendezvous needs both parties.
The sender-side edges are not optional: buffered sends never block (\textsc{Send} with room, \textsc{Loss} at capacity), but a synchronous send is a third blocking primitive, and a cycle that threads a sender-wait through receiver-waits is invisible to the receiver-side edges alone.
For example, $p_1 \triangleq c_1?x \then c_2!m \then \cdots$ and $p_2 \triangleq c_2?y \then c_1!n \then \cdots$, with $c_1$ asynchronous from $p_2$ to $p_1$ and $c_2$ synchronous from $p_1$ to $p_2$, deadlock: $p_1$ blocks on $c_1?x$ waiting for $p_2$, while $p_2$ blocks on $c_2?y$ waiting for $p_1$---a genuine deadlock surfaced by the Lean mechanisation of deadlock freedom over the full rule set.
On a platform that orders synchronous calls, as MicroKit does by priority (\textsc{Chan-Call}), cycles whose synchronous edges all point one way are excluded already; the mixed cycles are what acyclicity rules out.
(\textsc{Wf}2) rests on two shape requirements.
A standalone guard $[b]\,\rho$ would block for ever once $b$ is false, with no rule to rescue it, so guards occur only as branches.
Branches are \emph{event-headed}, of the shape $[b]\,\mathit{evt} \then \rho$ (nested choices of event-headed branches allowed), so that ``branches offering the same event'' is well-defined; this also excludes bare recursion-variable branches such as $\mu X.\,\alpha \then (X \extchoice \alpha \then X)$, which are genuinely non-deterministic after one unfolding---a hole surfaced by the Lean mechanisation of determinism.
Since \textsc{Guard-T} fires a guard and its event in one step, every branch offers its head event under its guard, which is what the labels $[b]\,\ell$ of $\Lambda^p$ record; disjointness then says that for a given event at most one branch is enabled, while branches offering distinct events impose no constraint, since the environment selects which event fires.
Exhaustiveness is required of plain choices only: a $\mathbf{default}$ choice is exhaustive by construction via $\rho_d$.
The last clause exists because \textsc{Ext-Default}'s negative premises ask whether the branches are stuck: a computation event or an asynchronous send is always enabled, so its stuckness is decided by the guard alone, whereas a receive depends on the buffer and a synchronous event on the rendezvous partner's state, and without the clause the escape could fire while a rendezvous stood ready, silently discarding it.
The clause constrains only the branches directly under a $\mathbf{default}$; the dispatch head of a body is a plain choice, so dispatch on several channels, including synchronous receives at the receive head of a body, is unaffected.

Both conditions can be checked statically, from the channel declarations and program text alone.
(\textsc{Wf}2) does two jobs: branch disjointness makes the fired event determine the branch, which is what determinism needs, and exhaustiveness keeps some branch of a plain choice enabled, which is what deadlock freedom needs.

\subsection{Preservation}\label{sec:preservation}

Throughout this section, a global configuration is \emph{reachable} if it is obtainable from the initial configuration $\mathcal{G}_0$ by a finite sequence of global transitions; the deadlock-freedom and determinism theorems are properties of this reachable state space, not of arbitrary configurations.
The invariants of the memory model, \eqref{inv:disj} and \eqref{inv:coh}, are static: no global transition changes a mapping, so they hold at every configuration once they hold at the declarations.
Coherence of the views, \eqref{inv:gcoh}, holds by construction, since a component's view is the restriction of $\Gamma$ read at the moment it steps.
What a transition can disturb is the channel store itself, and preservation is about that.

\begin{definition}[Buffer invariant]\label{def:bufwf}
A channel store $\Gamma$ satisfies $(\Gamma\textsc{-BufWf})$ when its domain is exactly the asynchronous channels, so a synchronous channel carries no buffer, and $|\Gamma(c)| \leq N_c$ for every asynchronous channel $c$.
\end{definition}

\begin{theorem}[Preservation]\label{thm:preservation}
If $\langle \mathbf{S},\, \Gamma \rangle \to_g \langle \mathbf{S}',\, \Gamma' \rangle$ and $\Gamma$ satisfies $(\Gamma\textsc{-BufWf})$, then so does $\Gamma'$.
Consequently every reachable configuration satisfies it, since $\mathcal{G}_0$ satisfies it.
\end{theorem}

\begin{proof}[Proof sketch]
By case analysis on the global rule.
\textsc{G-Sync} leaves $\Gamma$ unchanged.
\textsc{Ileave} writes back the stepping component's view, which differs from $\Gamma$ by at most one local channel step: a computation event or \textsc{Loss} leaves it unchanged; \textsc{Send} appends one value to $\Gamma(c)$ under the premise $|\Gamma(c)| < N_c$, so the bound is kept; \textsc{Recv} removes the head of $\Gamma(c)$, which only shortens it.
All three keep $\mathit{dom}(\Gamma)$, since they update an existing entry.
Along a run from $\mathcal{G}_0$ the invariant is carried by induction on the run's length; at $\mathcal{G}_0$ it holds by assumption, as a boot from empty buffers whose sends are \textsc{Send} or \textsc{Loss} steps guarantees.
\end{proof}

The theorem needs neither well-formedness nor reachability: it is a property of every transition.
Its use is in the theorems that follow, which read the kind of a channel off the store---a buffered channel is asynchronous, a declared channel with no buffer is synchronous---so that a component waiting on a channel is known to wait on the right kind of primitive.

\subsection{Deadlock Freedom}\label{sec:deadlock-free}

A component is never finished: when the residual of its invocation has terminated, its invocation is a fresh body, and a fresh body is receive-headed, so it is never terminated.
There is thus no terminated global configuration to exempt from the progress obligation; what remains to distinguish from deadlock is the idle system.
The environment is not modelled, so a receive on a channel whose sender is the environment---an interrupt---cannot be rendered as ``always enabled''.
The \emph{open-system assumption} is therefore stated in the conclusion of the theorem rather than as an environment transition rule, which would make the theorem trivially true whenever an interrupt channel exists.

\begin{definition}[Waiting on the environment]\label{def:waits-env}
A component $p$ \emph{waits on the environment} in $\langle \mathbf{S}, \Gamma \rangle$ when its invocation $\mathit{inv}(s^p)$ offers a receive on a declared channel $c$ at its enabled frontier, the sender of $c$ is not a component, and $\Gamma$ holds no message on $c$.
\end{definition}

The \emph{enabled frontier} of a process is where its next event is decided: through guards that hold, through the left operand of $\,;\,$ once terminated, and through the branches of a choice, by the same structural descent as enabledness.

\begin{theorem}[Deadlock Freedom]\label{thm:deadlock-freedom}
If the system is well-formed and has at least one component, then every reachable global configuration has an enabled global transition, or contains a component that waits on the environment.
\end{theorem}

\begin{corollary}[Progress of closed systems]\label{cor:progress}
If moreover every channel has a component as its sender, every reachable configuration has an enabled global transition.
\end{corollary}

\begin{proof}[Proof sketch]
The argument has four steps, which the mechanisation follows.
\emph{Blocked trichotomy.}
For a guarded process whose choices satisfy (\textsc{Wf}2), and each of whose send channels is either buffered within capacity or has no buffer entry, one of the following holds in a given state: it is enabled, it is terminated, it has a ready receive offer at its enabled frontier, or it holds a ready send on a bufferless channel, a blocked rendezvous sender; a process that is not enabled with a ready receive offer \emph{waits}, its buffer being empty.
Computation events, $\Omega$ and buffered sends are always enabled (\textsc{Send} or \textsc{Loss}), and a plain choice has an enabled branch by exhaustiveness, so the blocking primitives are an empty-buffer receive and a bufferless send; guarded recursion reaches the frontier in finitely many structural steps.
\emph{The invocation.}
Along every run the residual of a declared component keeps the structural well-formedness bundle (subject reduction), and so does its invocation, the residual or the fresh body once the residual has terminated (\textsc{MK-Restart}); an invocation is never terminated, as a fresh body is receive-headed.
\emph{Channel mentions.}
A residual mentions only channels its declared body mentions, so a wait on $c$ certifies that $c$ is declared, and the kind of $c$ is read off the buffer invariant (preservation): buffered means asynchronous, bufferless means synchronous.
\emph{Core.}
Suppose no global transition is enabled.
Then no invocation is enabled, since an enabled local step lifts to an \textsc{Ileave} step, and by the trichotomy every declared component waits: receive-side with no message, or at a synchronous send with no buffer.
If some receive-wait faces the environment, the second disjunct holds.
Otherwise every wait yields an edge of the (\textsc{Wf}1) dependency graph inside the declared components: a receive-wait points at the channel's declared sender, declared since the wait does not face the environment, and a synchronous send-wait at its declared receiver, declared by the endpoint facts of well-formedness.
A finite set in which every node has an outgoing edge contains a cycle, contradicting acyclicity.
The corollary follows because the second disjunct names an environment-sourced channel.
The full proof is the mechanised one.
\end{proof}

\paragraph{Remarks.}
The theorem rests on well-formedness, the initial configuration, the existence of one component, and reachability; the mechanisation carries the structural bundle of every residual and the buffer invariant along runs and discharges both at the initial configuration.
Fairness of the platform's scheduler is outside the formalism, and we do not assume it.

\subsection{Determinism}\label{sec:determinism}

We show that the global LTS $\zeta$ is \emph{deterministic}: for any reachable configuration and any label $\ell \in \Sigma_g$, there is at most one $\ell$-successor.
The \emph{glts} deliberately retains the nondeterminism inherent to concurrency---which component the (unmodelled) scheduler runs next, and when the environment raises an interrupt or delivers a message; the theorem states that this is the \emph{only} nondeterminism: it resides entirely in the choice \emph{of} the next label, never \emph{within} one.
By induction over a run this yields \emph{replay determinism}: the same sequence of labels reproduces the same configurations, a routine induction we do not mechanise.
It does not yield the stronger ``same inputs, same outputs'': which enabled label fires next is the scheduler's choice, so two schedules of the same external inputs may produce different output sequences; that stronger property would need every component to run an invocation to completion before another is scheduled, a run-to-completion assumption that is itself too strong for a system whose scheduler is outside the formalism.
Because computation events are labels of $\zeta$, the theorem covers them as well as communication: the same event fired from the same configuration has one effect.
Determinism is not a trace property and is not transferred by the refinement of the next section; it is proved of $\zeta$ directly.

\begin{theorem}[Determinism]\label{thm:determinism}
If the system is well-formed, then for every reachable global configuration $\langle \mathbf{S},\, \Gamma \rangle$ and every label $\ell \in \Sigma_g$, there is at most one configuration $\langle \mathbf{S}',\, \Gamma' \rangle$ such that $\langle \mathbf{S},\, \Gamma \rangle \xrightarrow{\ell} \langle \mathbf{S}',\, \Gamma' \rangle$.
\end{theorem}

\begin{proof}[Proof sketch]
Two ingredients: \emph{routing}, which pins the label to one stepping component or endpoint pair, and \emph{local determinism}, which pins that component's step.

\emph{Local determinism from (\textsc{Wf}2).}
A step of a term whose choices satisfy (\textsc{Wf}2) fires exactly one of the flattened branches of the choice it goes through: the branch's merged guard holds and its head is the step's event.
Two steps from one choice with the same underlying event, guard annotations aside, therefore reach the same branch, or two branches that disjointness makes exclusive; by induction on the derivation, two local steps with the same underlying event from one local state agree on the residual, the store and the view.
The ingredient applies to the invocation a reachable configuration gives a component, a fresh body or the residual of one, both of which carry the bundle (subject reduction); the default clause likewise survives steps and offers, so a residual's rendezvous offers are disciplined too, and a component outside the declarations stays at $\mathit{Skip}$ along every run, both discharged at $\mathcal{G}_0$ like the bundle.

\emph{Computation event $(p, \alpha)$.}
The label names the component; two \textsc{Ileave} steps with this label are two local steps of $p$'s invocation with the same underlying event, which agree by local determinism.

\emph{Async send $c!v$ and receive $c?v$.}
A channel-labelled local step starts from a term that mentions that channel, and a residual mentions only channels its declared body mentions, so the stepping component is the declared endpoint of $c$, unique by the declarations; a component outside the declarations sits at $\mathit{Skip}$ and has no step.
Within that component, local determinism applies; the sent value is $\llbracket e \rrbracket_{\sigma}$ of the active branch, and the received value is the head of $\Gamma(c)$.

\emph{Sync rendezvous $c.v$.}
\textsc{G-Sync} matches a send offer and a receive offer on $c$ at the enabled frontiers of the two endpoints' invocations.
Offers are determined like steps: at most one send offer and one receive offer on $c$ per term, by disjointness over the flattened branches together with the default clause of (\textsc{Wf}2), which keeps every synchronous event out of the branches of a $\mathbf{default}$ choice, so that a rendezvous passes such a choice only through its escape $\rho_d$ and never through \textsc{Ext-Default}'s negative premises; the endpoints are the declared ones, and $v$ is the sender's evaluated expression.

In all cases the transition is fixed by a unique component or endpoint pair, a unique branch, and a unique value, so at most one $\ell$-successor exists.
\end{proof}

\section{Abstraction: the Abstracted Local LTS}\label{sec:pd-global}

The glts carries every computation event of every component, so a property of the system's communication is checked over a state space that computation inflates.
The separation of computation from communication gives the way out.
A computation event changes only its component's local store $\sigma$ and never the global channel store $\Gamma$, so the local state changes between two channel events can be absorbed into one big step, and the abstracted components composed into a smaller model of the global behaviour.
This section defines that abstraction: each \emph{llts} is abstracted into an \emph{abstracted local LTS} (\emph{alts}) by \emph{hiding} its computation events---re-labelling each as the conventional silent event $\tau$, invisible to other components---and collapsing the resulting $\tau$-chains, so that only channel events and the guards that resolve choices remain visible.
Its purpose is the verification of channel-level trace properties, such as the diode's one-way flow, on the smaller model: the \emph{alts} of a component has exactly the channel traces of its \emph{llts}, and the composition of the \emph{alts}es exactly those of the glts, both proved below, so a trace property established on the abstracted global state space holds of the complete one, and a counterexample found there is a behaviour of the system.
Determinism, which is not a trace property, is proved on the glts itself.

As for the \emph{llts}, the figures draw an \emph{alts} as an STG, but the object defined and mechanised is a transition system over the same states as the \emph{llts}: no rule removes a state, and every rule adds a big-step edge between existing states.

So the \emph{alts} keeps every state: a big-step of $n$ computation steps between two communication events is one transition, but the states it spans remain, since each is a global interleaving point at which another component may step.

Write $\Lambda_c = \Sigma_c \cup \{\, [b]\,\ell \mid \ell \in \Sigma_c \,\}$ for the \emph{possibly-guarded channel labels}, ranged over by $\lambda$: \textsc{Guard-T} emits guarded channel events $[b]\,\ell$ directly (e.g.\ from every guarded receive branch of a choice), so the absorption rules must accept a guarded label wherever a bare channel event may appear.
Absorption is moreover \emph{canonical}, in two respects.
First, it is \emph{leading-only}: a $\tau$-chain is absorbed into the channel event that follows it, never into the one that precedes it, so a trailing chain becomes the leading chain of the \emph{next} event's big-step.
Second, an absorbed hidden step must be \emph{committed}---every move enabled at its source is silent and leads to the same state---written (cm) in the rules: \textbf{absorbing a hidden step never resolves a choice}.
A choice-resolving hidden step remains an explicit big-step $\btrans{\tau}$ (\textsc{Collapse}), which is where the choice visibly resolves.
Without canonicity one channel event would admit several big-steps differing only in the silent progress bundled with it, and a label of the composition would lead to several successors.
Table~\ref{tab:hide-collapse} gives the two re-labellings and the four big-step rules: \textsc{Hide}/\textsc{Hide-G} re-label a computation event as $\tau$/$[b]\tau$; \textsc{Collapse}/\textsc{Collapse-G} contract $\tau$-chains to big-steps; \textsc{Seq-Pre} absorbs a leading $\tau^*$-chain into the following channel label $\lambda \in \Lambda_c$; and \textsc{Seq-G} merges a guarded $[b]\tau$ with the following channel label via the guard merge; the view $\gamma$ of a state is $\Gamma$ by construction (G-Coh).
The result is a \emph{clean} \emph{alts} in which every transition is either a big-step (possibly guarded) $\tau$-chain $\btrans{\tau}$, or a channel label $\lambda \in \Lambda_c$ with its leading $\tau$-steps absorbed.

\begin{table}[t]
\scriptsize
\centering
\begin{tabular}{@{} p{0.47\linewidth} @{\quad} p{0.47\linewidth} @{}}
\begin{minipage}[t]{\linewidth}
\begin{mathpar}
\inferrule*[left=\textsc{Hide}]
  {\langle \rho,\, \sigma,\, \gamma \rangle \ltrans{\alpha} \langle \rho',\, \sigma',\, \gamma \rangle \\
    \alpha \in \Sigma_{\alpha} \quad \text{(G-Coh)}}
  {\langle \rho,\, \sigma,\, \Gamma \rangle \ltrans{\tau} \langle \rho',\, \sigma',\, \Gamma \rangle}
\end{mathpar}
\end{minipage}
&
\begin{minipage}[t]{\linewidth}
\begin{mathpar}
\inferrule*[left=\textsc{Hide-G}]
  {\langle \rho,\, \sigma,\, \gamma \rangle \ltrans{[b]\alpha} \langle \rho',\, \sigma',\, \gamma \rangle \\
    \alpha \in \Sigma_{\alpha} \quad \text{(G-Coh)}}
  {\langle \rho,\, \sigma,\, \Gamma \rangle \ltrans{[b]\tau} \langle \rho',\, \sigma',\, \Gamma \rangle}
\end{mathpar}
\end{minipage}
\\[0.5em]
\begin{minipage}[t]{\linewidth}
\begin{mathpar}
\inferrule*[left=\textsc{Collapse}]
  {\langle \rho,\, \sigma,\, \gamma \rangle \ltrans{\tau}^{+} \langle \rho',\, \sigma',\, \gamma \rangle \quad \text{(G-Coh)}}
  {\langle \rho,\, \sigma,\, \Gamma \rangle \btrans{\tau} \langle \rho',\, \sigma',\, \Gamma \rangle}
\end{mathpar}
\end{minipage}
&
\begin{minipage}[t]{\linewidth}
\begin{mathpar}
\inferrule*[left=\textsc{Collapse-G}]
  {\langle \rho,\, \sigma,\, \gamma \rangle \ltrans{[b]\tau} \langle \rho',\, \sigma',\, \gamma \rangle \\
    \langle \rho',\, \sigma',\, \gamma \rangle \ltrans{\tau}^{*} \langle \rho'',\, \sigma'',\, \gamma \rangle \quad \text{(G-Coh)}}
  {\langle \rho,\, \sigma,\, \Gamma \rangle \btrans{[b]\tau} \langle \rho'',\, \sigma'',\, \Gamma \rangle}
\end{mathpar}
\end{minipage}
\\[0.5em]
\begin{minipage}[t]{\linewidth}
\begin{mathpar}
\inferrule*[left=\textsc{Seq-Pre}]
  {\langle \rho,\, \sigma,\, \gamma \rangle \ltrans{\tau}^{*} \langle \rho',\, \sigma',\, \gamma \rangle \\
    \langle \rho',\, \sigma',\, \gamma \rangle \ltrans{\lambda} \langle \rho'',\, \sigma'',\, \gamma' \rangle \\
    \lambda \in \Lambda_c \quad \text{(G-Coh)} \quad \text{(cm)}}
  {\langle \rho,\, \sigma,\, \Gamma \rangle \btrans{\lambda} \langle \rho'',\, \sigma'',\, \Gamma' \rangle}
\end{mathpar}
\end{minipage}
&
\begin{minipage}[t]{\linewidth}
\begin{mathpar}
\inferrule*[left=\textsc{Seq-G}]
  {\langle \rho,\, \sigma,\, \gamma \rangle \ltrans{[b]\tau} \langle \rho',\, \sigma',\, \gamma \rangle \\
    \langle \rho',\, \sigma',\, \gamma \rangle \ltrans{\lambda} \langle \rho'',\, \sigma'',\, \gamma' \rangle \\
    \lambda \in \Lambda_c \quad \text{(G-Coh)} \quad \text{(cm)}}
  {\langle \rho,\, \sigma,\, \Gamma \rangle \btrans{[b]\cdot\lambda} \langle \rho'',\, \sigma'',\, \Gamma' \rangle}
\end{mathpar}
\end{minipage}
\end{tabular}
\caption{Abstraction rules.}
\label{tab:hide-collapse}
\end{table}

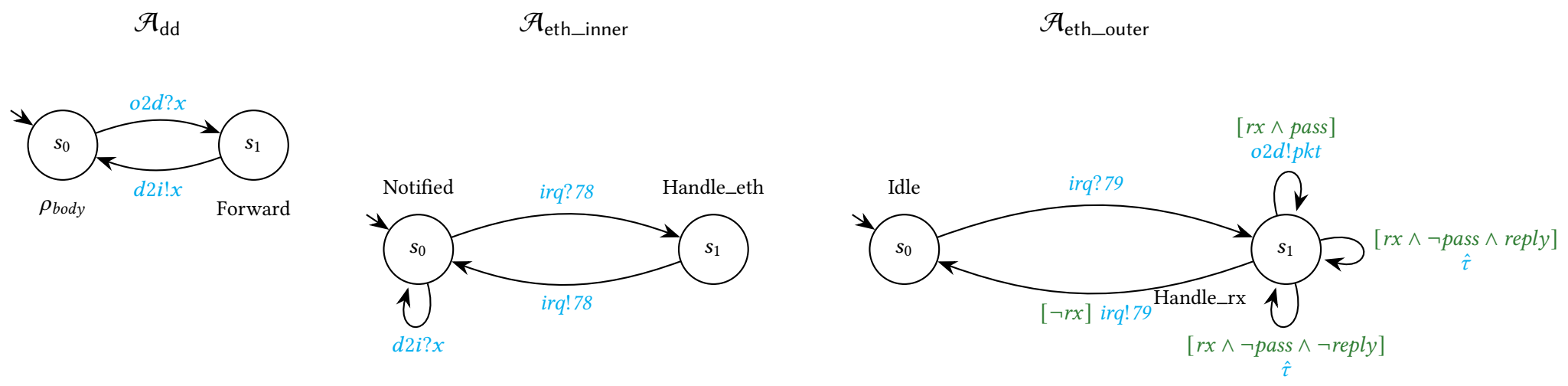
\begin{figure*}[t]
\centering
\begin{tikzpicture}[scale=0.75, transform shape,
  >=Stealth,
  every state/.style={minimum size=8mm, inner sep=1pt, font=\scriptsize},
  ev/.style={font=\scriptsize, text=cyan, align=center},
  stlbl/.style={font=\scriptsize, text=black}]

\begin{scope}[xshift=0.5cm]
\node[font=\small\bfseries] at (1.1, 1.4) {$\mathcal{A}_{\mathsf{dd}}$};
\node[state] (d0) at (0,   0) {$s_0$};
\node[state] (d1) at (2.2, 0) {$s_1$};
\draw[->] (-0.6,0.4) -- (d0);
\path[->]
  (d0) edge[bend left=20]  node[ev,above]{$o2d?x$}  (d1)
  (d1) edge[bend left=20]  node[ev,below]{$d2i!x$}  (d0);
\node[stlbl,below=3pt of d0]{$\rho_{\mathit{body}}$};
\node[stlbl,below=3pt of d1]{Forward};
\end{scope}

\begin{scope}[xshift=4.6cm]
\node[font=\small\bfseries] at (1.8, 1.4) {$\mathcal{A}_{\mathsf{eth\_inner}}$};
\node[state] (i0) at (0,  -1.2) {$s_0$};
\node[state] (i1) at (3.4, -1.2) {$s_1$};
\draw[->] (-0.6,-0.8) -- (i0);
\path[->]
  (i0) edge[bend left=20]  node[ev,above]{$\mathit{irq?78}$}          (i1)
  (i1) edge[bend left=20]  node[ev,below]{$\mathit{irq!78}$}          (i0)
  (i0) edge[loop below]    node[ev]{$d2i?x$}                          (i0);
\node[stlbl,above=3pt of i0]{Notified};
\node[stlbl,above=3pt of i1]{Handle\_eth};
\end{scope}

\begin{scope}[xshift=10.2cm]
\node[font=\small\bfseries] at (2.2, 1.4) {$\mathcal{A}_{\mathsf{eth\_outer}}$};
\node[state] (e0) at (0,  -1.2) {$s_0$};
\node[state] (e2) at (4.4, -1.2) {$s_1$};
\draw[->] (-0.6,-0.8) -- (e0);
\path[->]
  (e0) edge[bend left=20]  node[ev,above]{$\mathit{irq?79}$}                                                                       (e2)
  (e2) edge[bend left=20]  node[ev,below]{$\textcolor{guardgreen}{[\neg\mathit{rx}]}$\ $\mathit{irq!79}$}                          (e0)
  (e2) edge[loop above]    node[ev]{$\textcolor{guardgreen}{[\mathit{rx} \wedge \mathit{pass}]}$\\$o2d!\mathit{pkt}$}              (e2)
  (e2) edge[loop right]    node[ev]{$\textcolor{guardgreen}{[\mathit{rx} \wedge \neg\mathit{pass} \wedge \mathit{reply}]}$\\$\hat{\tau}$}      (e2)
  (e2) edge[loop below]    node[ev]{$\textcolor{guardgreen}{[\mathit{rx} \wedge \neg\mathit{pass} \wedge \neg\mathit{reply}]}$\\$\hat{\tau}$}  (e2);
\node[stlbl,above=3pt of e0]{Idle};
\node[stlbl,below left=2pt of e2]{Handle\_rx};
\end{scope}

\end{tikzpicture}
\caption{The abstracted STGs $\mathcal{A}_p$ of the diode components.}
\label{fig:ap-diode}
\end{figure*}

\paragraph{Example: Data Diode Abstracted LTS}
Applying the abstraction rules to the \emph{llts}es generated from the STGs of Figure~\ref{fig:llts-diode} yields the \emph{alts}es of Figure~\ref{fig:ap-diode}, again drawn as STGs: computation events are hidden, then absorbed into the following channel event where committed, or kept as explicit guarded big-steps $\hat{\tau}$ where they resolve a choice; interrupt notifications are channel events and are never hidden.
The drawing omits an interior state whose only incoming big-step is an absorbed chain and whose only outgoing one is the channel event that absorbed it, a convention of the figures and not a step of the abstraction, which keeps every state.
The diode computes nothing, so its \emph{alts} is its \emph{llts}. Two states the drawing keeps illustrate the two limits of absorption: a choice point (\textsf{eth\_outer}'s Handle\_rx) is where a hidden step would resolve a choice and so stays an explicit big-step, and a state between two channel events (\textsf{eth\_inner}'s Handle\_eth) has nothing hidden left to absorb---absorption fuses hidden steps into a channel event, never one channel event into another.

\subsection{Composing the Abstractions}\label{sec:composition}
The composition of the \emph{alts}es applies the two rules of Table~\ref{tab:sos-compose} to \emph{alts} big-steps of the invocations in place of \emph{llts} steps, a silent big-step being a silent step of the composition, and \textsc{G-Sync} verbatim, since it matches the process terms of the two endpoints, which the abstraction does not change.

\begin{figure*}[t]
\centering
\begin{tikzpicture}[scale=0.75, transform shape,
  >=Stealth,
  every state/.style={minimum size=10mm, inner sep=2pt, font=\small},
  ev/.style={font=\scriptsize, text=cyan, align=center},
  stlbl/.style={font=\scriptsize, text=black}
]

\node[state, initial above] (g0) at (0,   0)   {$g_0$};
\node[state]               (gi) at (-3.6, 0)  {$g_i$};
\node[state]               (g1) at (4.5, 0)   {$g_1$};
\node[state]               (g2) at (8.5, 0)   {$g_2$};

\path[->]
  (g0) edge[bend right=20]
    node[ev, above]{$\mathit{irq?78}$}                                                 (gi)
  (gi) edge[bend right=20]
    node[ev, below]{$\mathit{irq!78}$}                                                 (g0)

  (g0) edge
    node[ev, above]{$\mathit{irq?79}$}                                                 (g1)

  (g1) edge[bend left=35]
    node[ev, above, pos=0.5]{$\textcolor{guardgreen}{[\neg rx]}\ \mathit{irq!79}$}    (g0)

  (g1) edge[loop above]
    node[ev]{$\textcolor{guardgreen}{[b_2]}\ \hat{\tau}$}                              (g1)

  (g1) edge
    node[ev, above, pos=0.4]{$\textcolor{guardgreen}{[b_1]}\ o2d.v$}                   (g2)

  (g2) edge[bend left=30]
    node[ev, below]{$d2i.v$}                                                           (g1);

\node[stlbl, below=11pt of g0]{$\langle$\textsf{Idle, Notified, Notified}$\rangle$};
\node[stlbl, above=2pt of gi]{$\langle$\textsf{Idle, Notified, Handle\_eth}$\rangle$};
\node[stlbl, anchor=north west] at ([xshift=7pt,yshift=-8pt]g1.south){$\langle$\textsf{Handle\_rx, Notified, Notified}$\rangle$};
\node[stlbl, above=3pt of g2]{$\langle$\textsf{Handle\_rx, Forward, Notified}$\rangle$};

\end{tikzpicture}
\caption{Composition of the abstracted diode components (partial).}
\label{fig:global-lts}
\end{figure*}
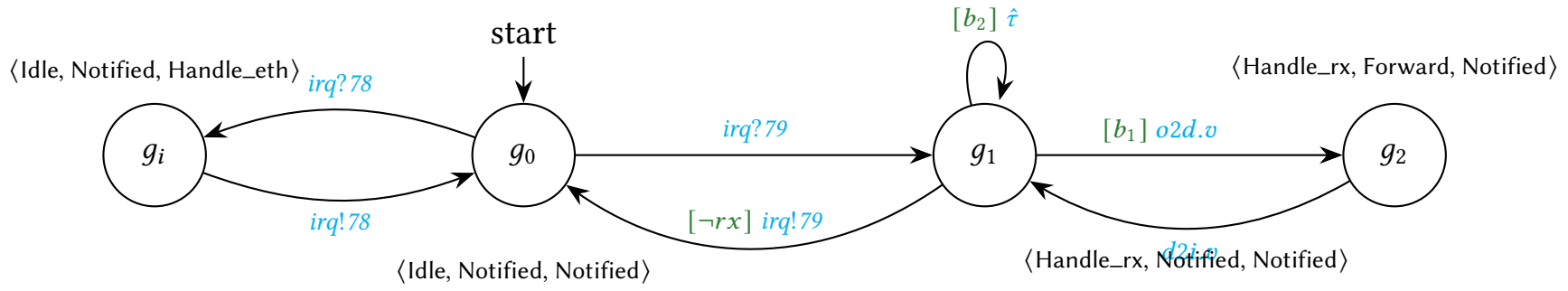

\paragraph{Example: Global Composition.}
Figure~\ref{fig:global-lts} shows the core forwarding behaviour of the composition of the \emph{alts}es of Figure~\ref{fig:ap-diode}, annotated with $\langle \textsf{eth\_outer}, \textsf{dd}, \textsf{eth\_inner} \rangle$ control states; its $\Sigma_c$-traces are those of $\zeta$, which admits further interleavings.
A label $c.v$ on an asynchronous channel abbreviates the matched pair $(c!v,\, c?v)$, two independent \textsc{Ileave} steps drawn as one edge; $b_1 \triangleq \mathit{rx} \wedge \mathit{pass}$ and $b_2 \triangleq \mathit{rx} \wedge \neg\mathit{pass}$.
No transition in $\zeta$ sends on $d2i$ from \textsf{eth\_outer} or receives on $o2d$ from \textsf{eth\_inner}:
the declarations (\textsc{Chan-Async}, \textsc{Chan-Sync}) ensure those transitions are absent.

\subsection{Refinement}\label{sec:alts-refinement}
The composition of the \emph{alts}es loses and adds no channel behaviour of the system; the result rests on one fact per component, that each \emph{alts} has the channel traces of its \emph{llts}, proved first.

The sequence of observable events along an execution is a \emph{trace}, and we compare processes by standard CSP \emph{trace refinement}~\cite{roscoe2010understanding}: $P \trefines Q$ when every observable trace of $Q$ is also a trace of $P$, so safety properties established over the traces of $P$ carry over to $Q$.

Hiding alone preserves traces: \textsc{Hide} and \textsc{Hide-G} re-label each computation step as a $\tau$-step and leave channel steps unchanged, so the $\Sigma_c$-traces of the hidden \emph{llts} $p^{\tau}$ are those of $p$.

\begin{lemma}[Refinement]\label{lem:refinement}
The $\Sigma_c$-traces of $\mathcal{A}_p$, obtained from $p^{\tau}$ by \textsc{Collapse}, \textsc{Collapse-G}, \textsc{Seq-Pre} and \textsc{Seq-G}, are exactly the $\Sigma_c$-traces of $p$; in particular $p \trefines \mathcal{A}_p$.
\end{lemma}

\begin{proof}[Proof sketch]
Each big-step of $\mathcal{A}_p$ unfolds to an \emph{llts} run with the same $\Sigma_c$-projection: \textsc{Collapse} and \textsc{Collapse-G} to a chain of hidden computation steps, which project to nothing, and \textsc{Seq-Pre} and \textsc{Seq-G} to the absorbed chain followed by the channel step, the guard merge leaving the projection unchanged.
Concatenating the unfoldings of a run of $\mathcal{A}_p$ gives a run of $p$ with the same $\Sigma_c$-trace, which is $p \trefines \mathcal{A}_p$; the converse inclusion is the next lemma.
\end{proof}

\begin{lemma}[Completeness]\label{lem:alts-completeness}
Every \emph{llts} step is a big-step of $\mathcal{A}_p$ at its hidden label, from the same state to the same state; consequently every finite $\Sigma_c$-trace of $p$ is a $\Sigma_c$-trace of $\mathcal{A}_p$.
\end{lemma}

\begin{proof}
Every \emph{llts} step has a big-step counterpart with the same $\Sigma_c$-projection: a hidden computation event becomes $\btrans{\tau}$ (\textsc{Collapse} with a one-step chain) or $\btrans{[b]\tau}$ (\textsc{Collapse-G} with an empty chain), and a channel label $\lambda \in \Lambda_c$ becomes $\btrans{\lambda}$ (\textsc{Seq-Pre} with zero absorbed steps).
Every state of the \emph{llts} is a state of the \emph{alts}, so the counterparts compose, replaying any \emph{llts} path big-step by big-step with an identical $\Sigma_c$-projection.
\end{proof}

\begin{theorem}[Composition refinement]\label{thm:composition-refinement}
For a well-formed system and a reachable configuration $\mathcal{G}$, the $\Sigma_c$-traces of the composition of the \emph{alts}es from $\mathcal{G}$ are exactly the $\Sigma_c$-traces of the glts $\zeta$ from $\mathcal{G}$; in particular $\zeta \;\trefines\; \Interleave_{p \in P}\, \mathcal{A}_p$.
The inclusion of the traces of $\zeta$ holds at every configuration, and from the initial configuration well-formedness alone suffices.
\end{theorem}

\begin{proof}[Proof sketch]
The inclusion of the glts needs no hypothesis: an \emph{llts} step is an \emph{alts} big-step at its hidden label (the single-step fact above), so every step of $\zeta$ is a step of the composition of the \emph{alts}es whose label is the $\Sigma_c$-projection of its own, a computation event projecting to a silent step.
For the converse, a big-step of the composition is realised by a non-empty \emph{llts} run of one component with the same $\Sigma_c$-projection (the refinement lemma), which lifts step by step through \textsc{Ileave} under that component: the first step reads the component's invocation, and every later state of the run is a residual that still steps and is therefore not terminated, so it is its own invocation.
That a stepping residual is not terminated holds of guarded terms, which is where well-formedness and reachability enter: the structural bundle that every residual of a well-formed system carries along runs (subject reduction), with undeclared components parked at $\mathit{Skip}$, supplies the guardedness.
A rendezvous is one step of both compositions.
\end{proof}

\section{Related Work}\label{sec:related}

\paragraph{Programs as models.}
Model checkers for process calculi and timed automata, SPIN~\cite{holzmann97spin}, FDR~\cite{gibson14fdr}, UPPAAL~\cite{uppaal} and PAT~\cite{sun09pat}, verify a model written beside the code, whose communication is by rendezvous or over queues that never drop and whose computation is absent or silent; CoreFi's program is the model, and its channels and computation are the platform's.
The P language~\cite{desai13p} makes the program the model: state machines exchanging events over unbounded queues, compiled to executable code and explored by a systematic tester, and ModP~\cite{desai18modp} composes such machines by assume-guarantee reasoning.
CoreFi shares the thesis and changes the abstraction: bounded and synchronous channels instead of unbounded queues, shared buffers instead of no shared memory, local code entered through named events rather than opaque foreign functions, and guarantees by construction rather than by exploration.
Ivy~\cite{padon16ivy} decides safety of parameterised protocols in a restricted logic, and code generation from verified specifications---PGo~\cite{finn23pgo,hackett25tracelink} from PlusCal and TLA+, Liu et al.~\cite{liu20maude} from Maude, AutoMan~\cite{sun25automan} with refinement proofs for hand optimisation---carries a verified design to code; in each, the communication model is the specification's, not the platform's.
Synchronous languages such as Lustre~\cite{halbwachs91lustre} and SCADE make the program the model for clocked control software, verified as written and compiled by qualified generators; their components communicate by synchronous dataflow within one clock, where CoreFi's run as separate components over buffers in shared memory.
BIP~\cite{basu06bip} builds systems from atomic components with ports, glued by connectors and coordinated by priorities, and is the closest component model to CoreFi's; its correctness-by-construction results concern the glue, whereas CoreFi's conditions concern the components' choices and their channel dependencies.

\paragraph{Deadlock freedom by typing.}
Multiparty session types~\cite{honda16multiparty} guarantee progress for well-typed sessions by projecting a global protocol onto its participants, over reliable, typically unbounded channels.
CoreFi's conditions are checked on the components as written, with no global protocol, and hold for bounded channels that drop; acyclicity of the wait relation and partition of choices replace the protocol's structure.

\paragraph{Embedded systems.}
MIMOS~\cite{yi20mimos} and Mimosa~\cite{huber25mimosa,huber25mimosa-compiler} give deterministic models and languages for embedded real-time software in which determinism follows from the communication discipline; CoreFi obtains determinism of the composed system, at most one successor for each label, from choice partition and keeps the platform's scheduling choice as its only nondeterminism.
Lingua Franca~\cite{lohstroh21lf} obtains determinism from logical timestamps on reactor events, a discipline over the coordination rather than the components' choices.
For seL4~\cite{klein09sel4}, Tower~\cite{hickey14tower} coordinates tasks written in the Ivory DSL over channels and was flown on seL4 in SMACCMPilot, Kalas~\cite{pohjola22kalas} compiles global programs to verified code with CakeML over unbounded channels, an idealisation CoreFi does not make, and Pancake~\cite{pancake} is a verified systems language for the local code CoreFi's computation events call.
The scheduler of seL4's mixed-criticality extension~\cite{lyons18scheduling} is what a timed version of the platform rule would model.

\section{Conclusion}\label{sec:conclusion}

We have presented CoreFi, a core calculus for embedded systems whose components compute on private state and communicate over bounded channels in shared memory.
Its local semantics has no silent transition; the platform that runs the components enters as one rule of the composition, the invocation, and an initial configuration, which is all an instantiation must supply, and we have supplied them for seL4 MicroKit.
Every system keeps its buffers within capacity, and two structural well-formedness conditions make every well-formed system deadlock-free in an open-system form and deterministic, each label of the composed system leading to at most one successor; an abstraction that hides computation has the channel traces of each component, and the composition of the abstractions those of the system.
The semantics, the platform rule, the theorems and the refinements are mechanised in Lean~4.

Three lines of work follow.
The first is the implementation: a compiler for the language that checks well-formedness and emits an executable image for seL4 MicroKit, and case studies of practical systems written in it beyond the diode.
The second is the correspondence between that implementation and the mechanisation, so that what runs is what the theorems speak about, in the manner of certified compilation.
The third is verification: properties beyond the structural ones, checked over the global state space or over the abstracted one that \S\ref{sec:pd-global} shows to have the same channel traces, and the question of how far either scales to the state spaces of practical embedded systems.
Last, seL4's mixed-criticality scheduler~\cite{lyons18scheduling}, which gives each component a budget and a period, is the natural next instance of the platform rule, bringing timing and fairness into the formalism.

\section*{Data-Availability Statement}
The artifact is the Lean~4 development that mechanises the semantics and the theorems of this paper; it is released as open source~\cite{fidus}.

\bibliographystyle{ACM-Reference-Format}
\bibliography{reference}

\end{document}